\documentclass[a4paper,11pt]{article}
\usepackage{jcappub}

\usepackage{lmodern}
\usepackage{empheq}
\usepackage{bm}
\usepackage[normalem]{ulem}
\usepackage{latexsym}
\usepackage{comment}
\usepackage{color}
\usepackage{dcolumn}
\usepackage{amsmath,amsfonts,amssymb}
\usepackage{graphicx,epsfig}
\usepackage{psfrag}
\usepackage{amsthm}
\usepackage{tabularx}
\newcolumntype{C}{>{\centering\arraybackslash}X}
\usepackage[toc,title,page]{appendix}

\title{Inflation is Not Magic}

\author[a,b]{S. Shajidul Haque,}
\author[c]{Ghadir Jafari,}
\author[a]{Bret Underwood}

\affiliation[a]{Department of Mathematics and Applied Mathematics,\\
University of Cape Town, South Africa}

\affiliation[b]{The National Institute for Theoretical and Computational Sciences, \\
Private Bag X1, Matieland, South Africa}

\affiliation[c]{Department of Physics Education\\ Farhangian University\\ P.O. Box 14665-889\\ Tehran, Iran}

\affiliation[d]{Department of Physics,\\
Pacific Lutheran University,\\
Tacoma, WA 98447, United States}

\emailAdd{bret.underwood@plu.edu}
\emailAdd{shajid.haque@uct.ac.za}

\abstract{
    Cosmological perturbations generated during inflation exhibit striking quantum features, including entanglement and large circuit complexity, yet their observational signatures remain effectively indistinguishable from classical stochastic variables. We explore this tension by describing inflationary perturbations as continuous-variable stabilizer states with vanishing quantum magic as measured by Wigner negativity, a necessary resource for universal quantum computation. We further show that the Wigner negativity arising from primordial non-Gaussianity is suppressed by the smallness of the non-linearity parameter $f_{NL}$ and the scalar power spectrum $\Delta_\zeta$. Identifying inflation as a generator of ``high complexity, low magic'' states provides an alternative information-theoretic perspective of the origin of structure in the universe.
}

\begin{document}

\maketitle
\flushbottom

\section{Introduction}

The theory of cosmological inflation 
\cite{Starobinsky:1980te,Guth:1980zm,Linde:1981mu,Albrecht:1982wi,Linde:1983gd} 
provides a compelling mechanism for generating the primordial curvature perturbations of the universe that seed large-scale structure (see \cite{Mukhanov}).
In this picture, quantum fluctuations of spacetime are stretched and amplified by the rapid expansion, producing pairs of entangled modes that are well described as two-mode squeezed vacuum states \cite{PhysRevD.42.3413,PhysRevD.50.4807}. 
These fluctuations exhibit several hallmarks of strongly quantum behavior.
As highly squeezed states they saturate the uncertainty principle, with their large amplitude uncertainty offset by a correspondingly small phase uncertainty.
Measures of non-classical correlations, such as entanglement entropy \cite{Brandenberger:1992sr,Brandenberger:1992jh,Brahma:2020zpk,Campo:2008ju,Campo:2008ij} and quantum discord \cite{Martin:2015qta,Martin:2021znx,Martin:2022kph}, grow with time.
In addition, the circuit complexity of inflationary perturbations increases linearly and saturates proposed bounds on complexity growth \cite{Bhattacharyya:2020rpy,Bhattacharyya:2020kgu,Haque:2021hyw,Haque:2024ldr,Haque:2021kdm}.

Despite these quantum signatures, cosmological perturbations exhibit strikingly classical behavior.
Long-wavelength modes can be accurately described by classical stochastic variables, largely independent of whether the modes have decohered \cite{Polarski:1995jg,Campo:2004sz,Kiefer:2008ku,Martin:2015qta,Micheli:2022tld,Nambu:2023nem}.
This tension raises a natural question: to what extent is the full machinery of quantum mechanics actually required to describe cosmological perturbations?
Quantum information theory offers a complementary perspective. The Gottesman–Knill theorem \cite{Gottesman:1998hu} shows that quantum circuits initialized in stabilizer states and acted upon solely by Clifford operations can be efficiently simulated on a classical computer, even though such circuits may feature superposition and entanglement.
A state that cannot be described within this stabilizer framework is said to possess quantum magic \cite{Bravyi:2004isx}, a necessary resource for universal quantum computation.

It is worth emphasizing that quantum magic constitutes an alternative notion of quantumness than squeezing, entanglement, or discord.
Indeed, a highly entangled two-mode squeezed state with non-vanishing quantum discord is nonetheless a Gaussian stabilizer state, and is thus classically simulable.
Moreover, entanglement entropy and quantum discord of super-horizon cosmological perturbations are  not directly observable quantities.
Quantum magic, as measured by Wigner negativity, is different in this respect: Hudson's theorem ties quantum magic to the non-Gaussianity of the primordial perturbations which are, in principle, observable.
Measurements of the bispectrum and higher-order correlators of the primordial curvature perturbations \cite{Planck:2019kim} thus make quantum magic an information-theoretic diagnostic with an observational handle.

In this paper, we reinterpret inflationary cosmological perturbations through the framework of stabilizer states and quantum magic of quantum information theory.
Although non-Gaussianity has been widely studied in cosmology, its interpretation as a quantum computational resource remains largely unexplored (though see the recent work \cite{Ireland:2026txt}).
Our main results are the following.
First, we show that Gaussian inflationary perturbations are continuous-variable stabilizer states with vanishing quantum magic.
Moving beyond the Gaussian sector, we show that the Wigner negativity sourced by a third-order cumulant is inherently non-perturbative.
The resulting closed-form Wigner function takes the form of a Gaussian times an Airy function, and oscillations in the Airy function produce an exponentially suppressed negativity ${\mathcal N} \lesssim \exp[-1/(6\gamma^2)]$ in terms of the dimensionless skewness $\gamma \ll 1$ of the distribution.
Finally, we study the skewness induced by the cosmological non-Gaussianity parameter $f_{NL}$, finding that current bounds on $f_{NL}\lesssim {\mathcal O}(10)$ imply that the resulting cosmological negativity is highly suppressed ${\mathcal N} \lesssim \exp[-10^6]$, while the negativity for slow-roll inflation is even more strongly suppressed.

Section \ref{sec:Stabilizer} reviews the stabilizer formalism, and we extend it to continuous-variables in Section \ref{sec:ContinuousStabilizer}.
Section \ref{sec:Cumulant} (together with Appendix \ref{app:Cumulant}) introduces the cumulant formulation of non-Gaussian Wigner functions, while  Appendix \ref{app:CubicGate} demonstrates that an explicit cubic phase gate gives a similar Airy structure and negativity estimate for the Wigner function.
Section \ref{sec:InflationaryPerts} formulates the dynamics of inflationary perturbations in the stabilizer language, and connects the non-Gaussianity of the bispectrum to the cumulant analysis of Section \ref{sec:Cumulant}.
Finally, in Section \ref{sec:Discussion} we conclude with implications of our results for decoherence the quantum-to-classical transition, and the prospects for observationally constraining the quantumness of the early universe through quantum magic.

\section{Stabilizers, Wigner Functions, and Cumulants}

\subsection{Stabilizers for Qubits}
\label{sec:Stabilizer}

Let's first start by reviewing the stabilizer formalism in the context of qubits.
Tracking the evolution of a generic $N$-qubit state
\begin{equation}
|\psi\rangle = \sum_i c_i |x_i\rangle\, ,
\end{equation}
the computational basis states $|x_i\rangle = |x_1 x_2\ldots x_N\rangle$ are built from tensor products of single qubit basis states $\{|0\rangle,|1\rangle\}^{\otimes N}$,
requires 
${\mathcal O}(2^N)$ complex amplitudes $\{c_i\}$ and is therefore intractable for large $N$.
However, not all physically interesting evolutions explore the full Hilbert space. 
The stabilizer formalism identifies a family of states and operations still capable of creating entanglement yet whose evolution remains efficiently simulable.
Let ${\mathcal P}_N$ denote the $N$-qubit Pauli group generated by tensor products of the Pauli operators $\{\hat I, \hat X, \hat Y,\hat Z\}$ with phases $\{\pm 1, \pm i\}$.
An element of the Pauli group can be written as a Pauli string such as $\hat X_1 \hat Y_2\ldots\hat Z_N$, where subscripts indicate the qubit on which each operator acts.
A {\bf stabilizer state} $|\psi\rangle$ is the unique simultaneous $+1$ eigenstate of $N$ independent,  commuting Pauli operators $\{\hat S_1,\ldots S_N\}\subseteq {\mathcal P_N}$ (excluding $-\hat I$)
\begin{equation}
    \hat S_i |\psi\rangle = |\psi\rangle\,.
\end{equation}
These operators generate an Abelian subgroup of ${\mathcal P}_N$, known as the {\bf stabilizer group}.
Stabilizer states need not be product states and can include entanglement: the Bell state $|\psi\rangle = (|00\rangle + |11\rangle)/\sqrt{2}$ is stabilized by the generators $\hat X_1\otimes \hat X_2$ and $\hat Z_1 \otimes Z_2$ (the operator $-\hat Y_1 \otimes \hat Y_2$ is also in the stabilizer group as it is generated by the product of the generators).

A generic unitary $\hat U$ maps stabilizer generators to non-Pauli operators $\hat S_i \rightarrow \hat U \hat S_i \hat U^\dagger\notin {\mathcal P}_N$, making the stabilizer description impractical for arbitrary dynamics. However, there exists a restricted set of unitaries known as the {\bf Clifford group} ${\mathcal C}_N$, that leave the Pauli group fixed under conjugation, e.g.~$\hat U \hat P \hat U^\dagger \in {\mathcal P}_N$ for $\hat P \in {\mathcal P}_N$ and $\hat U \in {\mathcal C}_N$.
The Clifford group is generated by familiar gates such as the Hadamard ($\hat H$), phase ($\hat S$), and controlled-NOT (CNOT).
For example, the Pauli string $\hat X_1 \otimes \hat {\mathbb I}$ transforms under the Clifford group elements as
\begin{align}
   \hat H (\hat X_1 \otimes \hat {\mathbb I}) \hat H^\dagger &= \hat Z_1 \otimes \hat {\mathbb I}\,; \\
   \hat S (\hat X_1 \otimes \hat {\mathbb I}) \hat S^\dagger &= \hat Y_1 \otimes \hat {\mathbb I}\,; \\
   {\rm CNOT}\, (\hat X_1 \otimes \hat {\mathbb I}) {\rm CNOT}^\dagger &= \hat X_1 \otimes \hat X_2\,.
\end{align}
Conjugation by operators in the Clifford group map stabilizer generators $\{\hat S_i\}$ to a new set of transformed stabilizer generators $\hat S'_i = \hat U \hat S_i \hat U^\dagger$ that stabilize the new state
$|\psi'\rangle = \hat U |\psi\rangle$ since
\begin{align}
    |\psi'\rangle = \hat U |\psi\rangle &= \hat U \hat S_i |\psi\rangle \\
    &= \left(\hat U \hat S_i \hat U^\dagger\right) \hat U |\psi\rangle\\
    & = \hat S' |\psi'\rangle\,.
\end{align}
Thus, the evolution of a stabilizer state under a Clifford circuit is obtained by only updating the $N$ stabilizer generators rather than tracking $2^N$ complex amplitudes.
This is the essence of the {\bf Gottesman-Knill Theorem} \cite{Gottesman:1998hu,Aaronson:2004xuh}, which guarantees that quantum circuits composed solely of Clifford operations, Pauli measurements, and computational-basis state preparations can be simulated efficiently on a classical computer.
Stabilizer states also play a role in quantum error correction, where error-detecting code spaces arise as the $+1$ eigenspace of stabilizer states.

To achieve a computational advantage over classical simulation a quantum circuit must therefore utilize some non-Clifford operations \cite{Bravyi:2004isx}. 
A quantity that tracks the ``non-stabilizerness" of a quantum state is called {\bf magic}. 
A useful \emph{magic monotone} for an $N$-qubit circuit is the negativity of the (discrete) Wigner function \cite{Veitch:2012ttw,Emerson:2013zse,Pashayan:2015cos,Mari:2012ypq}
\begin{equation}
    {\mathcal N}_{N} = \frac{1}{2} \sum_{q,p} \left(|W_\rho(q,p)| - 1\right)\,.
\end{equation}
The discrete Wigner function $W_\rho(q,p)$ is constructed using a basis of phase-point operators $\hat A(q,p)$
\begin{equation}
    W_\rho(q,p) = \frac{1}{N} {\rm Tr}\left[\hat \rho \hat A(q,p)\right]\,,
\end{equation}
where the phase space $(q,p)$ consists of a set of $2^{2N}$ discrete points. The negativity of a Wigner function constructed from stabilizer states should vanish \cite{Gross:2006wkl}, establishing negativity as a measure of the quantum magic of the state.

In the next section, we will generalize the stabilizer formalism to the continuous variable case.

\subsection{Stabilizers in Continuous Variable Systems}
\label{sec:ContinuousStabilizer}

We now generalize the stabilizer formalism to continuous variables (CV) by
defining our computational basis using the Fock (number) basis $\{|n\rangle\}_{n=0}^\infty$.
Unlike standard CV approaches like GKP \cite{Gottesman:2000di} that utilize position eigenstates\footnote{For an exception, see \cite{Sanders:2002xod}.}
\cite{Braunstein:1997db,Lloyd:1997up,Bartlett:2002vmm,Bartlett:2002geb,Mari:2012ypq,Weedbrook:2011wxo}, the number basis is more natural for discussing particle creation in cosmology.
See Table \ref{table:CVStabilizer} for a comparison of our CV stabilizer formalism to that of the GKP CV stabilizer approach \cite{Gottesman:2000di}.
In this formulation, we 
construct a co-moving stabilizer framework. Rather than utilizing the Heisenberg-Weyl group as the Pauli group, we utilize the quadratic operators
\begin{equation}
    \hat K_+ = \frac{1}{2}(\hat a^\dagger)^2, \quad \hat K_- = \frac{1}{2}\hat a^2, \quad \hat K_0 = \frac{1}{2}\hat a^\dagger \hat a \,,
\end{equation}
where $\hat a$ and $\hat a^\dagger$ are annihilation and creation operators satisfying $[\hat a, \hat a^\dagger] = 1$, and we removed the zero-point fluctuation from $\hat K_0$. These generators operate within the universal enveloping algebra of ${\rm su}(1,1)$, satisfying the commutation relations
\begin{equation}
    [\hat K_0, \hat K_\pm] = \pm \hat K_\pm, \quad [\hat K_-, \hat K_+] = 2\hat K_0 + \frac{1}{2}\,,
\end{equation}
where the offset in the algebra is due to the removal of the zero-point term from $\hat K_0$.
The exponential of these generators forms the squeezing operator $\hat S(\xi) = \exp(\xi \hat K_+ - \xi^* \hat K_-)$ with $\xi = r e^{2i\phi}$ in terms of the squeezing parameter $r\in [0,\infty)$ and squeezing angle $\phi\in[-\pi/2,\pi/2]$, and the rotation operator $\hat R(\theta) = \exp(-i \theta \hat K_0)$ in terms of the rotation angle $\theta \in [0,2\pi]$, generating the group SU(1,1).

\renewcommand{\arraystretch}{1.3}
\begin{table}[t]
\centering
\footnotesize
\begin{tabularx}{1.0\textwidth}{|C||C|C|C|C|}
\hline
Stabilizer Formalism & Computational Basis & Pauli Group Analog & Clifford Group Analog & Stabilizer State Example\\ [0.5ex] \hline\hline
GKP \cite{Gottesman:2000di} & Position Eigenstates  & Heisenberg-Weyl & Jacobi Group &  Comb of position eigenstates \\
& $|q\rangle$ & & & $\sum_{s=-\infty}^\infty |q=\alpha s\rangle$ \\ [0.5ex] \hline
This paper & Number Eigenstates  & SU(1,1) & Jacobi Group & (Displaced) Squeezed States \\
& $|n\rangle$ & & & $|\alpha,\xi\rangle$ \\ [0.5ex] \hline
\end{tabularx}
\caption{A comparison of the GKP \cite{Gottesman:2000di} CV stabilizer formalism with the CV stabilizer formalism we will use in this work.}
\label{table:CVStabilizer}
\end{table}

The vacuum state $|0\rangle$ is the unique $+1$ eigenstate of the rotation operator 
\begin{equation}
    \hat R(\theta) |0\rangle = |0\rangle\,.
    \label{eq:CVStabilizer}
\end{equation}
Thus, we identify the stabilizer group as the $U(1)$ Abelian group generated by $\hat R(\theta)$.
To track the evolution of the vacuum under general Gaussian unitaries, we embed this stabilizer group within
the Jacobi group, defined as the semi-direct product of the symplectic group $SU(1,1)$ and the Heisenberg-Weyl group $\mathcal{H}$; the latter are generated by the displacement operators
\begin{equation}
    \hat D(\alpha) = \exp(\alpha \hat a^\dagger - \alpha^* \hat a)\,.
\end{equation}
The Jacobi group thus contains all Gaussian unitaries (rotations, squeezers, and displacements) generated by quadratic Hamiltonians. 
The evolution of the vacuum under a general element $\hat U$ of this ``CV Clifford" group yields a displaced squeezed vacuum state
\begin{equation}
    |\psi\rangle = |\alpha,\xi\rangle = \hat D(\alpha) \hat S(\xi) |0\rangle\,.
\end{equation}
Rather than defining a rigid code space, we utilize the stabilizer formalism dynamically to establish a co-moving stabilizer framework.
This allows us to track the evolving state by updating the stabilizer generators.
The new state $|\psi\rangle$ is stabilized by the conjugated rotation operator $\hat R'(\theta) = \hat U \hat R(\theta) \hat U^\dagger$. 
For example, under a 
Gaussian unitary
consisting of a displacement and squeezing operator $\hat U = \hat D(\alpha)\hat S(\xi)$, the rotation operator transforms as:
\begin{equation}
    \hat R'(\theta) = \hat D(\alpha) \hat S(\xi)\ \hat R(\theta)\ \hat S^\dagger(\xi)\hat D^\dagger(\alpha) = \exp\left(-i\theta \hat {\mathcal K}_0\right)\,,
\end{equation}
where the transformed
generator $\hat {\mathcal K}_0 = \frac{1}{2}\hat {\mathcal A}^{\dagger}\hat {\mathcal A}$ is written in terms of the Bogoliubov-transformed creation and annihilation operators
\begin{align}
    \hat {\mathcal A} &= \hat D(\alpha) \hat S(\xi) \hat a \hat S^\dagger(\xi)\hat D^\dagger(\alpha) = (\hat a-\alpha) \cosh r + (\hat a^\dagger - \alpha^*) e^{2i\phi} \sinh r\,; \\
    \hat {\mathcal A}^{\dagger} &= \hat D(\alpha) \hat S(\xi) \hat a^\dagger \hat S^\dagger(\xi)\hat D^\dagger(\alpha) = (\hat a^\dagger-\alpha^*) \cosh r + (\hat a - \alpha) e^{-2i\phi} \sinh r\,.
\end{align}
This transformed generator continues to define a $U(1)$ subgroup that uniquely characterizes the evolved Gaussian state through $\hat R'(\theta) |\alpha,\xi\rangle = |\alpha,\xi\rangle$.
Thus, we can follow the continuous
evolution of any stabilizer state $|\alpha,\xi\rangle$ under 
general Gaussian unitaries simply by tracking the conjugation
of the stabilizer group generated by $\hat R(\theta)$.

Even though the Hilbert space is infinite-dimensional, the evolution of stabilizer states under arbitrary Clifford group operations is restricted to Gaussian states $|\Psi\rangle_G$.
As in the discrete case, for continuous variables we will use ``magic'' in the sense of non-stabilizerness \cite{Bravyi:2004isx}; we quantify this by the negativity of the Wigner function, which has been used as an indicator for non-classicality \cite{Kenfack:2004ges,Mari:2012ypq,Albarelli:2018ujl,Takagi:2018rqp}.
In the continuous phase space $(q, p)$, the Wigner function for a Gaussian state $|\Psi\rangle_G$ is 
\begin{equation}
    W(q,p) = \frac{1}{\pi} \int_{-\infty}^{\infty} dy \, \langle q-y | \Psi\rangle_G\ {}_G\langle \Psi| q+y \rangle e^{2ipy}\, . 
    \label{eq:WignerDefinition}
\end{equation}
For single-mode Gaussian states, the Wigner function can be written as the Gaussian
\begin{equation}
    W_{G}(q, p) = \frac{1}{\pi} \exp\left( -2 \left[ \sigma_p^2 q^2 - 2 \sigma_{qp}\, q\, p + \sigma_q^2 p^2 \right] \right)\,,
    \label{eq:Wigner1}
\end{equation}
or equivalently, in more compact notation,
\begin{equation}
    W_G(\vec z) = \frac{1}{\pi} \exp\left(-\frac{1}{2} \vec z^T \mathbf{\Sigma}^{-1} \vec z \right)\, ,
    \label{eq:Wigner2}
\end{equation}
where $\vec z = (q, p)^T$ is the vector of phase-space coordinates
and $\mathbf{\Sigma}$ is the covariance matrix
\begin{equation}
    \mathbf{\Sigma} = \begin{pmatrix}
        \sigma_q^2 & \sigma_{qp} \cr
        \sigma_{qp} & \sigma_p^2
    \end{pmatrix}\, .
\label{eq:CovarianceMatrix}
\end{equation}
For a squeezed (but not displaced) vacuum state of a free harmonic oscillator with mass $m$ and frequency $\omega$, the variances are
\begin{align}
\label{eq:SqueezedVariance1}
    \sigma_q^2 &=\langle \hat q^2\rangle = \frac{1}{2m\omega} (\cosh(2r) - \sinh(2r)\cos2\phi)\,; \\
\label{eq:SqueezedVariance2}
    \sigma_p^2 &= \langle \hat p^2\rangle =\frac{m\omega}{2} (\cosh(2r) + \sinh(2r)\cos2\phi)\,; \\
\label{eq:SqueezedVariance3}
    \sigma_{qp} &=\frac{1}{2} \langle \hat q \hat p + \hat p \hat q\rangle = \frac{\hbar}{2} \sinh(2r)\sin2\phi\,.
\end{align}
The Wigner function describes an ellipse in phase space whose orientation and eccentricity are determined by the squeezing parameter $\xi = re^{2i\phi}$.

Hudson's Theorem \cite{HUDSON1974249} guarantees that for pure states, the Wigner function is non-negative ($W(q,p) \ge 0$) if and only if the state is Gaussian. Thus, non-Gaussianity is a pre-requisite for quantum magic.
Since our stabilizer states are Gaussian, they therefore possess vanishing Wigner negativity:
\begin{equation}
    {\mathcal N} = \int dq\, dp\, |W(q,p)| - 1 = 0\,.
    \label{eq:Negativity}
\end{equation}
In practice, to calculate the Wigner negativity we separate the Wigner function into its positive and negative regions
\begin{equation}
    W(q,p) = \begin{cases} [W]_+ > 0 \cr [W]_- < 0\, . \end{cases}
\end{equation}
Because the Wigner function is normalized, we have
\begin{align}
    &\int dq\ dp\ W(q,p) = \int dq\ dp\ [W]_+ + \int dq\ dp\ [W]_- = 1 \,, \nonumber \\
    & \Rightarrow \int dq\ dp\ [W]_+ = 1 - \int dq\ dp\ [W]_-\, .
\end{align}
Using this, the negativity \eqref{eq:Negativity} can be calculated by the volume of phase space where the Wigner function is negative
\begin{equation}
    {\mathcal N} = \int dq\ dp\ |W(q,p)| -1= \int dq\ dp\ [W]_+ - \int dq\ dp\ [W]_- -1= - 2\int dq\ dp\ [W]_-\, .
    \label{eq:Negativity2}
\end{equation}
Since Hudson's theorem applies to pure states, throughout this paper we will restrict our analysis to the pure states; it would be interesting to study how negativity can arise in non-Gaussian mixed states.

Another measure of magic is Mana, defined as \cite{Emerson:2013zse,Takagi:2018rqp,Albarelli:2018ujl}
\begin{equation}
    \text{Mana} = \log\left( \int dq\ dp\ |W(q,p)|\right)\, .
\end{equation}
The quantity Mana has the advantage that it is additive when considering additional modes for separable Wigner functions, since
\begin{equation}
    \text{Mana} = \log\left( \int d^nq\ d^np\ |W(\vec z)|\right) = \log \left(\prod_{i=1}^N \int dq_i\ dp_i\ |W(q_i,p_i)|\right) = \sum_{i=1}^N \log\left(\int dq_i\ dp_i\ |W(q_i,p_i)|\right)\, .
    \label{eq:Mana}
\end{equation}
in terms of the $2N$-component phase space vector $\vec z = (q_1,...q_N,p_1...p_N)$.

\subsection{Cumulant Expansion and Negativity}
\label{sec:Cumulant}

As established in the previous section, achieving a quantum computational advantage in the continuous variable domain requires non-Gaussian circuit elements to generate non-zero magic.
Consider, for instance, a unitary operator generated by a non-quadratic interaction Hamiltonian $\hat {\mathcal U}_{\rm non-Cliff} = e^{-i\hat H_{\rm int}t}$
where $\hat H_{\rm int}$ includes cubic terms like $(\hat a^\dagger)^2 \hat a$.
For a multi-mode system of $N$ harmonic oscillators with raising and lowering operators $\{\hat a_i,\hat a_i^\dagger\}$, such interactions take the form
\begin{equation}
    \hat H_{\rm int} = \sum_{i,j} \left[g_{ij} (\hat a_i^\dagger)^2 \hat a_j + g_{ij}^* \hat a_j^\dagger \hat a_i^2\right]\, .
\end{equation}
Under this evolution, the stabilizer \eqref{eq:CVStabilizer} for the $j$th mode $\hat R_j(\theta)$ transforms as 
\begin{equation}
\hat R_j(\theta) \rightarrow e^{i\hat H_{\rm int} t} \hat R_j(\theta) e^{-i\hat H_{\rm int} t} = e^{-i\frac{1}{2}\theta\ \hat O_j^\dagger \hat O_j}\, ,
\label{eq:NonCliffordStabilizerEvolution}
\end{equation}
where the time evolved operator $\hat a_j \rightarrow \hat O_j(t) = e^{i\hat H_{\rm int} t} \hat a_j e^{-i\hat H_{\rm int} t}$ is given by the Baker-Campbell-Hausdorff expansion
\begin{equation}
    \hat O_j(t) = \sum_{n=0}^\infty \frac{(it)^n}{n!} \underbrace{[\hat H_{\rm int},[\hat H_{\rm int},\ldots,[\hat H_{\rm int},\hat a_j]\ldots]]}_{n \text{ nested commutators}}\, .
    \label{eq:NonCliffordOperator}
\end{equation}
Unlike Gaussian operations, where the algebra closes, commutators with this cubic Hamiltonian $\hat H_{\rm int}$  generate higher-order polynomials in the creation and annihilation operators.
For example, the first-order commutator yields
\begin{equation}
    [\hat a_j, \hat H_{\rm int}] = \sum_{k} \left[2g_{jk}\ \hat a_{j}^\dagger \hat a_k + g^*_{kj} \hat a_k^2\right]\, .
\end{equation}
In general, nested commutators at level $n$ will produce ${\mathcal O}\left((2N)^{n}\right)$  higher-order operators. This phenomenon of operator spreading, where a simple initial operator evolves into an increasingly complex superposition of operators, is formalized by the framework of Krylov complexity (or K-complexity) \cite{Muck:2022xfc}. Within this framework, the expansion order $n$ tracks the operator's delocalization over its dynamically generated Krylov basis \cite{Parker:2018yvk}.
As a result, the generator of the transformed stabilizer \eqref{eq:NonCliffordStabilizerEvolution} no longer resides within the $SU(1,1)$ algebra.
The stabilizer group is therefore not preserved, and tracking the full state evolution under such non-Clifford interactions becomes computationally intractable, with complexity growing with the expansion order $n$.
This breaking of the stabilizer group structure leads to operator spreading and is a mechanism for the generation of magic in continuous variable systems.

A convenient tool to quantify the deviations from magic due to CV non-Gaussian interactions is the \emph{cumulant expansion}
of the generating function $\chi(\vec\xi)$
\begin{equation}
    W(\vec z) = \frac{1}{(2\pi)^{2N}} \int d^{2N} \xi\, e^{-i \vec{\xi}\cdot \vec{z}}\, \chi\left(\vec \xi\right)\, ,
    \label{eq:WignerFourierTransform}
\end{equation}
in terms of the $2N$-component Fourier transformed variables $\xi^a$.
The logarithm of the generating function admits the expansion
\begin{equation}
    \log \chi\left(\vec \xi\right) = \sum_{n\geq 1} \frac{i^n}{n!} C_{a_1\ldots a_n} \xi^{a_1}\ldots\xi^{a_n}\, ,
    \label{eq:CumulantGeneratingFunction}
\end{equation}
written as a sum over the {\bf connected cumulants} $C_{a_1\ldots a_n} = \langle z_{a_1}\ldots z_{a_n}\rangle_c$ of the phase space variables.
Connected cumulants measure the shape of a distribution relative to a Gaussian, so that for Gaussian states all cumulants with $n \geq 3$ vanish.
For example, for a single mode $N=1$ the first cumulant of the position coordinate is identical to the first moment, $\langle \hat q\rangle_c = \langle \hat q\rangle$. 
The second cumulant is the variance $\langle \hat q^2\rangle_c = \langle \hat q^2\rangle - \langle \hat q\rangle^2$ while
the third cumulant is found by subtracting the parts determined by the first and second moments,
\begin{equation}
    \langle \hat q^3\rangle_c = \langle \hat q^3\rangle - 3\langle \hat q^2\rangle\langle \hat q\rangle + 2 \langle \hat q\rangle^3\, .
\end{equation}
The fourth cumulant can be found by subtracting out the part that arises from the Gaussian,
\begin{equation}
    \langle \hat q^4\rangle_c = \langle \hat q^4\rangle - 3 \left(\langle \hat q^2\rangle_c\right)^2\, .
\end{equation}
In the case of vanishing first moments $\langle \hat q \rangle = 0$, the corresponding cumulants simplify to
\begin{equation}
    \langle \hat q\rangle_c = 0, \hspace{.2in} \langle \hat q^2\rangle_c = \langle \hat q^2\rangle, \hspace{.2in} \langle \hat q^3\rangle_c = \langle \hat q^3\rangle, \hspace{.2in} \langle \hat q^4\rangle_c = \langle \hat q^4\rangle - 3 \left(\langle \hat q^2\rangle\right)^2\, ,
\end{equation}
with similar expressions for cumulants involving $\hat p$ variables.

The Fourier transform \eqref{eq:WignerFourierTransform} with the generating function \eqref{eq:CumulantGeneratingFunction} gives a representation of the Wigner function generated from the Gaussian part
\begin{equation}
    W(\vec z) = \exp\left(\sum_{n\geq3} \frac{(-1)^n}{n!} C_{a_1...a_n}\partial_{a_1}\ldots\partial_{a_n}\right)W_G(\vec z)\, ,
    \label{eq:WignerExponential}
\end{equation}
where the derivatives act on the Gaussian Wigner function $W_G$ \eqref{eq:Wigner2}.
Expanding this expression for the first few terms linear in the $\{C_{a_1\ldots a_n}\}$ beyond the Gaussian
\begin{align}
    W(\vec z) &= W_G(\vec z) - \frac{1}{3!}C_{abc} \partial_a \partial_b \partial_c W_G(\vec z) + \frac{1}{4!}C_{abcd} \partial_a\partial_b \partial_c \partial_d W_G(\vec z) + \ldots
   \\
    &= W_G(\vec z) \left[1+\frac{1}{3!} C_{abc} H_{abc}(\vec z; \mathbf{\Sigma}) + \frac{1}{4!} C_{abcd} H_{abcd}(\vec z; \mathbf{\Sigma}) + \ldots\right]\, ,
    \label{eq:WignerCorrected}
\end{align}
where repeated indices are summed, $\partial_a \equiv \partial/\partial z^a$, and in the last line we rewrote the result in terms of the multivariate Hermite polynomials
\begin{equation}
    H_{a_1\ldots a_n}(\vec z, \mathbf{\Sigma}) = (-1)^n\, e^{\frac{1}{2}\vec z^T\mathbf{\Sigma}^{-1}\vec z}\, \partial_{a_1}\ldots\partial_{a_n} \left(e^{-\frac{1}{2}\vec z^T\mathbf{\Sigma}^{-1}\vec z}\right)\,.
    \label{eq:MultivariateHermite}
\end{equation}
The detailed form of the third-order and fourth-order cumulant expansions can be found in Appendix \ref{app:Cumulant}.

The cumulant-expanded Wigner function \eqref{eq:WignerCorrected} simplifies when the non-Gaussianity
lies along a principal axis of the underlying Gaussian. 
In terms of the dimensionless canonical variables $\tilde q=\sqrt{m\omega}\,q$, $\tilde p=p/\sqrt{m\omega}$, the covariance matrix \eqref{eq:CovarianceMatrix} has eigenvalues
\begin{equation}
    \lambda_\pm=\frac{\tilde\sigma_q^2+\tilde\sigma_p^2}{2}
    \pm\frac12\sqrt{(\tilde\sigma_q^2-\tilde\sigma_p^2)^2+4\tilde\sigma_{qp}^2}\, ,
    \label{eq:PrincipalVariances}
\end{equation}
which for the squeezed vacuum \eqref{eq:SqueezedVariance1}--\eqref{eq:SqueezedVariance3} are
$\lambda_\pm=\tfrac12e^{\pm2r}$. 
The rotation to the principal axes,
\begin{equation}
    X=\sin\phi\,\tilde q+\cos\phi\,\tilde p,\qquad
    Y=-\cos\phi\,\tilde q+\sin\phi\,\tilde p,
    \label{eq:PrincipalAxes}
\end{equation}
is a symplectic (Gaussian) transformation, so that $(X,Y)$ are canonical variables with diagonal covariance ${\rm diag}(\lambda_+,\lambda_-)$.
For simplicity, we will isolate the effect of skewness along the major axis by assuming that the only non-vanishing higher-order cumulant is the third cumulant along the major axis $C_{XXX} \neq 0$.
In terms of the scaled coordinates
$x=X/\sqrt{\lambda_+}$, $y=Y/\sqrt{\lambda_-}$ and the dimensionless skewness
\begin{equation}
    \gamma\equiv\frac{C_{XXX}}{\lambda_+^{3/2}} = \frac{\langle XXX\rangle_c}{\langle XX\rangle^{3/2}}\, ,
    \label{eq:GammaDef}
\end{equation}
the first-order correction \eqref{eq:WignerCorrected} is
\begin{equation}
    W(x,y)=W_G(x,y)\left[1+\frac{\gamma}{6}\left(x^3-3x\right)\right],
    \label{eq:W3CorrectionXY}
\end{equation}
where
\begin{equation}
    W_G(x,y)=\frac{e^{-x^2/2}}{\sqrt{2\pi}}\frac{e^{-y^2/2}}{\sqrt{2\pi}}\, ,
\end{equation}
is a unit-variance gaussian in the scaled $(x,y)$ coordinates.

At first glance, the cubic terms in \eqref{eq:W3CorrectionXY} source Wigner negativity when the term in square brackets becomes negative, for $x<0$ once $|x| \gtrsim (6/\gamma)^{1/3}$.
However, for a small deviation from Gaussianity $\gamma \ll 1$ this lies far out on the tail, $x_* = (6/\gamma)^{1/3}\gg 1$, where the cumulant expansion \eqref{eq:WignerCorrected} fails.
Specifically, \eqref{eq:W3CorrectionXY} keeps only the first term in a power series expansion of the exponential \eqref{eq:WignerExponential} in powers of $C_{XXX}$.
Far out on the tail these terms scale as $\sim \frac{1}{n!}(x/x_*)^{3n}$, so for $|x| \geq x_*$ these terms are all of the same order and cannot be truncated.

Instead of performing a perturbative expansion, when $C_{XXX}$ is the only non-zero higher-order cumulant the generating function \eqref{eq:CumulantGeneratingFunction} becomes
\begin{equation}
    \chi(\xi_X,\xi_Y)=\exp\left[-\frac12\lambda_+\xi_X^2-\frac12\lambda_-\xi_Y^2-\frac{i}{6}C_{XXX}\,\xi_X^3\right].
    \label{eq:TruncatedGeneratingFunction}
\end{equation}
The Wigner function \eqref{eq:WignerFourierTransform} now takes the form of a Gaussian along the minor axis times an Airy profile along the major axis,
\begin{equation}
    W(x,y)=\frac{e^{-y^2/2}}{\sqrt{2\pi}}\;{\mathcal A}_\gamma(x)\, ,
    \label{eq:WignerAiryForm}
\end{equation}
where
\begin{equation}
    {\mathcal A}_{\gamma}(x) = \left(\frac{2}{\gamma}\right)^{1/3} e^{u/\gamma + 1/(3\gamma^2)}\ {\rm Ai}\left(\left(\frac{2}{\gamma}\right)^{1/3} \left(x+\frac{1}{2\gamma}\right)\right) ,
    \label{eq:ScaledAiryFunction}
\end{equation}
with ${\rm Ai}(x)$ an Airy function of the first kind.
The profile is normalized, $\int {\mathcal A}_\gamma\, du = 1$, and reduces to a Gaussian as $\gamma \to 0$.
In the near-Gaussian limit $\gamma \ll 1$ expanding the Airy function leads to the first term in the Hermite series \eqref{eq:WignerCorrected}
\begin{equation}
    W(x,y) \approx W_G(x,y) \left[1+\frac{\gamma}{6} H_3\left(x\right) + \ldots\right],
\end{equation}
matching the correction \eqref{eq:W3CorrectionXY}.
However, unlike \eqref{eq:W3CorrectionXY}, the full Wigner function \eqref{eq:WignerAiryForm} does not suffer from the same issues with truncating the power series.

\begin{figure}[t]
\centering\includegraphics[width=\textwidth]{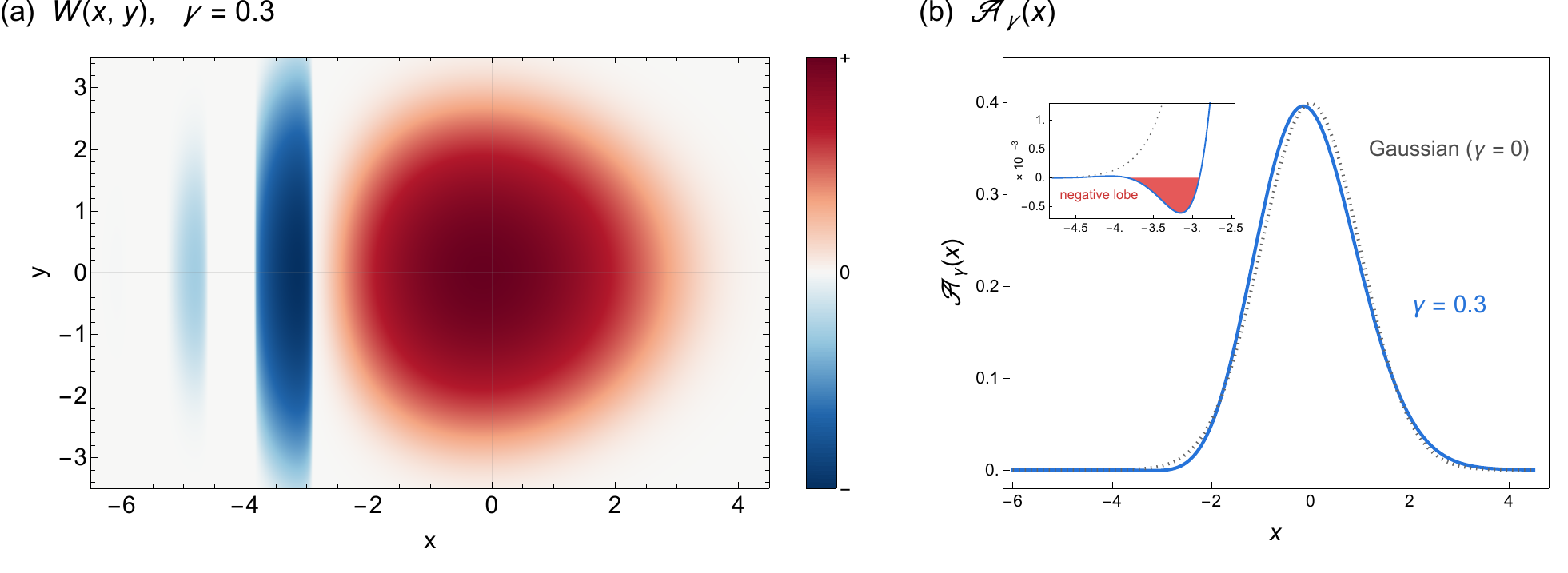}
\caption{(a) The non-Gaussian Wigner function \eqref{eq:WignerAiryForm} arising from the generating function \eqref{eq:TruncatedGeneratingFunction} with non-vanishing third-order cumulant $C_{XXX}\neq 0$ in terms of the scaled variables $x =X/\sqrt{\lambda_+},\ y = Y/\sqrt{\lambda_-}$ for $\gamma = 0.3$. The Wigner function has negative strips arising from the oscillations of the Airy function. (b) The negativity of the Wigner function is dominated by the first negative lobe of the scaled Airy function ${\mathcal A}_{\gamma}(u)$ \eqref{eq:ScaledAiryFunction}, shown here for $\gamma = 0.3$.}
\label{fig:AiryWigner}
\end{figure}

Working with the full Wigner function \eqref{eq:WignerAiryForm}, the Wigner negativity comes from the oscillations of the Airy function, as shown in Figure \ref{fig:AiryWigner}.
In particular, the zeros of ${\mathcal A}_\gamma(x)$ are located at $\tilde x_k = -1/(2\gamma) + (\gamma/2)^{1/3} a_k$ where the $a_i < 0$ are the traditional zeros of the Airy function, e.g.~$a_1 \approx -2.34, a_2 \approx -4.09,\ldots$.
Writing $x = -1/(2\gamma) + (\gamma/2)^{1/3} \zeta$, the scaled Airy function \eqref{eq:ScaledAiryFunction} becomes
\begin{equation}
    {\mathcal A}_{\gamma}(x) = \left(\frac{2}{\gamma}\right)^{1/3} \exp\left[-\frac{1}{6\gamma^2} + \frac{\zeta}{2^{1/3} \gamma^{2/3}}\right]\ {\rm Ai}(\zeta)\, .
\end{equation}
There is a strong suppression by $e^{-1/(6\gamma^2)} \ll 1$, and because of the strong dependence on $\zeta/\gamma^{2/3}$ the negativity is dominated by the  rightmost part of the first negative lobe of the Airy function.
Taylor-expanding about the zero ${\rm Ai}(a_1 + \delta) \approx {\rm Ai}'(a_1) [\delta + (a_1/6) \delta^3 + \ldots]$, the resulting negativity for $\gamma \ll 1$ becomes
\begin{align}
    {\mathcal N} &= -2\int dq\ dp\ \left[W(q,p)\right]_-\  = -2 \int dx\ dy\ \frac{e^{-y^2/2}}{\sqrt{2\pi}}\  \left[{\mathcal A}_{\gamma}\left(x\right)\right]_- = -2 \int dx\ \left[{\mathcal A}_\gamma(x)\right]_- \\
    & \approx 2e^{-1/(6\gamma^2)} \left|\int_{a_2}^{a_1} e^{\zeta/(2^{1/3}\gamma^{2/3})} {\rm Ai}(\zeta)\ d\zeta \right| \\
    &\approx  2^{5/3} \gamma^{4/3}\ |{\rm Ai}'(a_1)|\ \exp\left[-\frac{1}{6\gamma^2} - \frac{|a_1|}{2^{1/3} \gamma^{2/3}}\right] \left[1-2^{2/3} |a_1| \gamma^{4/3} + {\mathcal O}{(\gamma^2})\right]\, .
    \label{eq:NegativityAiryWigner}
\end{align}
The negativity arising from the third order cumulant parameterized through $\gamma = C_{XXX}/\lambda_+^{3/2} \ll 1$ is thus strongly suppressed by the non-perturbative terms in the exponential; for $\gamma = 0.05$ we find ${\mathcal N} \approx 5\times 10^{-37}$.

The simplified generating function \eqref{eq:TruncatedGeneratingFunction} is useful for estimating the effects of a non-Gaussian skew from $\langle XXX\rangle$ on the Wigner function and negativity; however, a true state must also have higher-order cumulants as well.
In Appendix \ref{app:CubicGate} we compute the Wigner function of a Gaussian state acted on by a cubic phase gate $e^{i\lambda \hat p^3}$, which sources $\langle XXX\rangle$. Its Wigner function is the same Airy profile evaluated along parabolic rather than straight curves in phase space,  and its negativity is identical in form to \eqref{eq:NegativityAiryWigner} with $\gamma = 12 \lambda\sigma_p^3$ (when $\sigma_{qp} = 0$), supporting the robustness of the suppression found in \eqref{eq:NegativityAiryWigner} and confirming that it is not an artifact of a simplified generating function.

\section{Inflationary Perturbations and Magic}
\label{sec:InflationaryPerts}

Having established the relation between stabilizer states, Gaussianity, and Wigner negativity in Section \ref{sec:ContinuousStabilizer}, we now apply this framework to cosmological perturbations generated during inflation in the early universe. 
Our goal in this section is to show that the standard quadratic theory of inflationary perturbations evolves entirely within the set of continuous-variable stabilizer states. 
As a result, despite producing large squeezing and significant entanglement, the quantum state contains no quantum magic.
We begin by reviewing the standard quantization of scalar perturbations, emphasizing that the dynamics is governed by an SU(1,1) algebra, and therefore restricted to Gaussian states.
We then reinterpret the two-mode squeezed vacuum as a CV stabilizer state whose stabilizer generators transform under inflationary evolution exactly as in the continuous-variable stabilizer formalism of Section \ref{sec:ContinuousStabilizer}.

We consider scalar perturbations of a spatially flat Friedmann-Lemaître-Robertson-Walker (FLRW) background metric during inflation
\begin{equation}
    ds^2 = -dt^2 + a(t)^2 d\vec x^2\, .
\end{equation}
The expansion rate of the background is characterized by the Hubble rate $H = \dot a/a$.
We will primarily be interested in accelerated expansion for which the Hubble rate is approximately constant, expressed in terms of the ``slow roll" parameters $\epsilon = -\dot H/H^2 \ll 1$ and $\eta = \dot \epsilon/(H\epsilon)$ with $|\eta| \ll 1$.
In the absence of anisotropic stress the metric perturbations
\begin{equation}
    ds^2 = a(\eta)^2\left(-(1+2\psi(\eta,\vec x))\, d\eta^2 + (1-2\psi(\eta,\vec{x}))\, d\vec x^2\right)\,.
\end{equation}
and perturbations of the inflaton scalar field $\varphi(\eta,\vec x) = \varphi_0(\eta) + \delta \varphi(\eta,\vec x)$ combine with the metric into the (gauge-invariant) curvature perturbation $\zeta = \psi + (H/\dot \varphi)\delta \varphi$.
Transforming to the canonically normalized Mukhanov curvature perturbation variable 
$v(\eta,\vec x) = M_{pl}a\sqrt{2\epsilon}\ \zeta = a(\eta) M_{pl} (\delta \varphi + \sqrt{2\epsilon}\, \psi)$ \cite{Mukhanov:1988jd}, the quadratic action is \cite{Mukhanov}
\begin{equation}
    S_2 = \frac{1}{2} \int d\eta d^3x \left[ v'^2 - (\partial_i v)^2 + \frac{z''}{z} v^2 \right],
\end{equation}
where a prime denotes a derivative with respect to conformal time ${\mathcal H}' = a'/a$ and where $z \equiv aM_{pl}\sqrt{2\epsilon}$.
Promoting the perturbation to a quantum field and expanding in Fourier modes
\begin{equation}
    \hat v(\eta,\vec x) = \int \frac{d^3k}{(2\pi)^{3/2}} \hat v_{\mathbf k}(\eta) e^{i\mathbf k\cdot \mathbf x}\,,
\end{equation}
we can define the operator and its canonical momentum as
\begin{equation}
    \hat v_{\mathbf k} = \frac{1}{\sqrt{2k}}\left(\hat c_{\mathbf k} + \hat c_{-\mathbf k}^\dagger\right)\, , \qquad \hat p_{\mathbf k} = -i\sqrt{\frac{k}{2}} \left(\hat c_{\mathbf k} - \hat c_{-\mathbf k}^\dagger\right)\,,
\end{equation}
where $\hat c_{\mathbf k},\hat c_{\mathbf k}^\dagger$ are creation and annihilation operators in the Heisenberg representation 
satisfying $[\hat c_{\mathbf k},\hat c_{\mathbf p}^\dagger] = \delta^3(\mathbf k-\mathbf p)$.
Since $\hat v(\eta,\vec{x})$ is real we have $\hat v_{\mathbf k}^\dagger = \hat v_{-\mathbf k}$.
Because the background is isotropic, the modes decouple into pairs with opposite momenta $(\mathbf{k}, -\mathbf{k})$. The Hamiltonian for each mode pair is
\begin{equation}
    \hat{H} = \int d^3k\, \hat {\mathcal H}_{\mathbf{k},-\mathbf{k}} = \int d^3k \left[ k \left( \hat{c}_{\mathbf{k}}^\dagger \hat{c}_{\mathbf{k}} + \hat{c}_{-\mathbf{k}}^\dagger \hat{c}_{-\mathbf{k}} + 1 \right) - i \frac{z'}{z} \left( \hat{c}_{\mathbf{k}} \hat{c}_{-\mathbf{k}} - \hat{c}_{\mathbf{k}}^\dagger \hat{c}_{-\mathbf{k}}^\dagger \right) \right]\, .
\label{eq:CosmoHamiltonian}
\end{equation}
The first term represents the free evolution of the modes with energy $k$, while the second term, proportional to the background expansion rate $z'/z(\eta)$, is a time-dependent interaction term that creates pairs of $(\mathbf{k},-\mathbf{k})$ particles from the vacuum due to the expanding background.
Crucially, this Hamiltonian is strictly quadratic in the creation and annihilation operators.
This immediately implies that if the Universe begins in a Gaussian state, it will remain Gaussian for all time regardless of the magnitude of squeezing or particle production.
Because the Hamiltonian \eqref{eq:CosmoHamiltonian} of cosmological perturbations does not mix modes of different $\mathbf k$, the Hilbert space factorizes into a continuous tensor product of independent two-mode $(\mathbf k,-\mathbf k)$ sectors.

The two-mode Bunch-Davies vacuum state is defined as the state annihilated by $\hat c_{\mathbf k}$ in the far past $|k\,\eta_i| \gg 1$
\begin{equation}
    \hat c_{\mathbf k}(\eta_i) |0_{\mathbf k},0_{-\mathbf k}\rangle = 0\, .
    \label{eq:BunchDavies}
\end{equation}
The unitary time-evolution of the vacuum state 
\begin{equation}
\hat {\mathcal U}_{\mathbf k}(\eta) = {\mathcal P} e^{-i\int_{\eta_i}^\eta \hat {\mathcal H}_{\mathbf{k},-\mathbf{k}}(\tilde \eta)d\tilde \eta}    
\end{equation}
can be factorized as a rotation and squeezing operation \cite{PhysRevD.42.3413,PhysRevD.50.4807}
\begin{equation}
    \hat {\mathcal U}_{\mathbf k} = \hat {\mathcal S}_{\mathbf k}(r_k,\phi_k)\hat {\mathcal R}_{\mathbf k}(\theta_k)\, ,
    \label{eq:CosmoUnitary}
\end{equation}
where $\hat {\mathcal R}_{\mathbf k}(\theta_k)$ is the two-mode rotation operator
\begin{equation}
    \hat {\mathcal R}_{\mathbf k}(\theta_k) \equiv \exp\left[-i\theta_k(\eta) (\hat c_{\mathbf k}(\eta_i)\hat c_{\mathbf k}^\dagger(\eta_i) + \hat c_{-\mathbf k}^\dagger(\eta_i)\hat c_{-\mathbf k}(\eta_i))\right]\, ,
\end{equation}
and $\hat {\mathcal S}_{\mathbf k}(r_k,\phi_k)$ is the two-mode squeezing operator
\begin{equation}
    \hat {\mathcal S}_{\mathbf k}(r_k,\phi_k) \equiv \exp\left[\frac{r_k(\eta)}{2}\left(e^{-2i\phi_k(\eta)}\hat c_{\mathbf k}(\eta_i)\hat c_{-\mathbf k}(\eta_i) - e^{2i\phi_k(\eta)}\hat c_{-\mathbf k}^\dagger(\eta_i)\hat c_{\mathbf k}^\dagger(\eta_i)\right)\right]\, .
\end{equation}
The vacuum state evolves through \eqref{eq:CosmoUnitary} into the two-mode squeezed state 
\begin{equation}
    |\Psi(\eta)\rangle_{\mathbf k,-\mathbf k} = \hat {\mathcal U}_{\mathbf k}(\eta) |0_{\mathbf k},0_{-\mathbf k}\rangle = \frac{1}{\cosh r_k}\sum_{n=0}^\infty (-1)^n e^{2in\phi_k(\eta)} \tanh^n \left(r_k(\eta)\right) |n_{\mathbf k},n_{-\mathbf k}\rangle\, .
    \label{eq:TwoModeSqueezedState}
\end{equation}
The non-zero two-point functions for the amplitude and momentum of the two-mode squeezed state are
\begin{align}
    \langle \hat v_{\mathbf{k}}\hat v_{\mathbf{k}'}\rangle
&=\frac{1}{2k}\!\left[\cosh2r_k-\cos(2\phi_k)\sinh2r_k\right](2\pi)^3\delta^3(\mathbf{k}+\mathbf{k}')\, ; \label{eq:vvTwoPoint}\\
\langle \hat p_{\mathbf{k}}\hat p_{\mathbf{k}'}\rangle
&=\frac{k}{2}\!\left[\cosh2r_k+\cos(2\phi_k)\sinh2r_k\right] (2\pi)^3\delta^3(\mathbf{k}+\mathbf{k}')\, ; \label{eq:ppTwoPoint} \\
\frac12\langle\{\hat v_{\mathbf{k}},\hat p_{\mathbf{k}'}\}\rangle
&=\frac{1}{2}\sinh2r_k\sin2\phi_k\, (2\pi)^3\delta^3(\mathbf{k}+\mathbf{k}')\, .
\label{eq:vpTwoPoint}
\end{align}
The time-dependence for the squeezing parameter $r_k(\eta)$, squeezing angle $\phi_k(\eta)$, and rotation angle $\theta_k(\eta)$ are determined by the equations of motion
\begin{align}
    \frac{dr_k}{d\eta} &= -\frac{z'}{z} \cos(2\phi_k)\, ; \\
    \frac{d\phi_k}{d\eta} &= -k + \frac{z'}{z} \coth(2r_k)\sin(2\phi_k)\, ; \\
    \frac{d\theta_k}{d\eta} &= k - \frac{z'}{z} \tanh(r_k) \sin(2\phi_k)\, .
\end{align}

During inflation, the background expansion is approximately de Sitter with scale factor $a(\eta) = -1/(H_{dS}\, \eta)$ for $-\infty < \eta < 0$. 
The squeezing and rotation parameters during inflation have the solution \cite{PhysRevD.50.4807}
\begin{align}
    \label{eq:dSSolution1}
    r_k &= \sinh^{-1} \left(\frac{1}{2k|\eta|}\right)\,; \\
    \label{eq:dSSolution2}
    \phi_k &= -\frac{\pi}{4} - \frac{1}{2} \tan^{-1} \left(\frac{1}{2k|\eta|}\right)\,; \\
    \label{eq:dSSolution3}
    \theta_k &= k|\eta| - \tan^{-1}\left(\frac{1}{2k|\eta|}\right)\,.
\end{align}
At early times $k|\eta| \gg 1$ the modes are inside the horizon and the squeezing is small $r_k \ll 1$, so that the state \eqref{eq:TwoModeSqueezedState} becomes the vacuum state \eqref{eq:BunchDavies}. 
At late times $k|\eta|\ll 1$ the modes exit the horizon the squeezing grows as the number of e-folds $r_k \sim \ln a \gg 1$, and the squeezing and rotation angles approach a constant $\phi_k \approx -\pi/2, \theta_k \approx -\pi/2$.
The two-point functions \eqref{eq:vvTwoPoint} - \eqref{eq:vpTwoPoint} become
\begin{align}
     \langle \hat v_{\mathbf{k}}\hat v_{\mathbf{k}'}\rangle
&=\frac{e^{2r_k}}{2k}(2\pi)^3\delta^3(\mathbf{k}+\mathbf{k}')\, ; \label{eq:vvTwoPointdS}\\
\langle \hat p_{\mathbf{k}}\hat p_{\mathbf{k}'}\rangle
&=\frac{k}{2}(2\pi)^3\delta^3(\mathbf{k}+\mathbf{k}')\, ; \label{eq:ppTwoPointdS}\\
\frac12\langle\{\hat v_{\mathbf{k}},\hat p_{\mathbf{k}'}\}\rangle
&=-\frac{e^{r_k}}{2} (2\pi)^3\delta^3(\mathbf{k}+\mathbf{k}')\, .
\label{eq:vpTwoPointdS}
\end{align}
Despite the extreme squeezing, the momentum variance is fixed at the vacuum value, and is exponentially smaller than the corresponding amplitude variance.
The corresponding two-point function for the amplitude of the co-moving curvature perturbation $\zeta_{\mathbf k}$ becomes
\begin{align}
    \langle \hat\zeta_{\mathbf k'} \hat\zeta_{\mathbf k}\rangle = \frac{\langle v_{\mathbf k'}v_{\mathbf k}\rangle}{2M_{pl}^2 a^2(\eta)\epsilon} \approx \frac{e^{2r_k}}{4 k a^2(\eta) M_{pl}^2\epsilon} (2\pi)^3 \delta^3(\mathbf k+\mathbf k') = \frac{2\pi^2}{k^3} \Delta_\zeta^2\ (2\pi)^3 \delta^3(\mathbf k + \mathbf k')\,,
    \label{eq:SqueezedTwoPointCurvature}
\end{align}
where $\Delta^2_\zeta = H_{dS}^2/(8\pi^2M_{pl}^2 \epsilon)\sim 2\times 10^{-9}$ is the small (constant) dimensionless power spectrum of the curvature perturbations.

The state $|\Psi(\eta)\rangle_{\mathbf k,-\mathbf k}$ \eqref{eq:TwoModeSqueezedState} describing cosmological perturbations during inflation is, in many respects, a strongly quantum state.
For example, the perturbations \eqref{eq:vvTwoPoint}-\eqref{eq:vpTwoPoint} saturate the Heisenberg uncertainty principle $\langle \hat v_{\mathbf k}\hat v_{\mathbf k'}\rangle \langle \hat p_{\mathbf k}\hat p_{\mathbf k'}\rangle - \langle\hat v_{\mathbf k}\hat p_{\mathbf k}\rangle^2 = 1/4$.
At late times when the modes are well outside the horizon the two-mode squeezed state \eqref{eq:TwoModeSqueezedState} is a {\bf highly entangled} EPR state
\begin{equation}
    |\Psi\rangle \sim \frac{1}{\cosh r_k} \sum_{n=0}^\infty |n_{\mathbf k},n_{-\mathbf k}\rangle\, .
\end{equation}
One way to quantify the amount of entanglement is with the entanglement entropy obtained by tracing out half of the degrees of freedom \cite{Brandenberger:1992sr,Brandenberger:1992jh,Brahma:2020zpk,Campo:2008ju,Campo:2008ij}
\begin{align}
    S_{ent} &= -\mbox{Tr}\left(\hat \rho_{\mathbf k} \ln \rho_{\mathbf k}\right) = (1+\sinh^2 r_k) \ln (1+\sinh^2 r_k) - \sinh^2 r_k\ \ln (\sinh^2 r_k) \nonumber \\
    & \approx \ln (\sinh^2 r_k) \approx 2r_k\,,
    \label{eq:entanglementEntropy}
\end{align}
where $\rho_{\mathbf k} = \mbox{Tr}_{-\mathbf k}\left(|\Psi\rangle_{\mathbf k -\mathbf k}\langle \Psi|_{\mathbf k,-\mathbf k}\right)$ and we used the approximation $r_k \gg 1$.
We see that the entanglement grows with the squeezing, and thus grows over time with the number of e-folds.
Alternatively, we can quantify the quantum-ness of this highly entangled state with \emph{quantum discord} \cite{Henderson:2001wrr,Ollivier:2001fdq}, which is the difference between the total information (as measured by the von Neumann entropy ${\mathcal I}(\mathbf k, -\mathbf k)$) and the classical mutual information ${\mathcal J}(\mathbf k,-\mathbf k)$, ${\mathcal D}(\mathbf k,-\mathbf k) = {\mathcal I}(\mathbf k,-\mathbf k) - {\mathcal J}(\mathbf k,-\mathbf k)$. 
In the classical limit, these two quantities are identical and the quantum discord vanishes.
As discussed in \cite{Martin:2015qta}, the quantum discord for the state \eqref{eq:TwoModeSqueezedState} grows with squeezing
${\mathcal D}(\mathbf k,-\mathbf k) \approx \frac{2}{\ln 2} r_k$ similarly to the entanglement entropy \eqref{eq:entanglementEntropy}.
Through these measures we see that inflationary perturbations are non-classical, at least in this sense.
Another way to study the quantum properties of the unitary evolution \eqref{eq:CosmoUnitary} and the corresponding state \eqref{eq:TwoModeSqueezedState} is through its quantum circuit complexity ${\mathcal C}(k)$, defined as the minimum number of fundamental unitary gates that are required to construct the given unitary evolution.
The circuit complexity thus measures the total resources needed to prepare the state, in contrast to quantum magic which measures only the resources needed beyond Gaussian operations \cite{Mari:2012ypq,Emerson:2013zse,Albarelli:2018ujl,Takagi:2018rqp}.
For cosmological squeezed states produced during inflation, the quantum circuit complexity, like the quantum discord, grows with the squeezing parameter ${\mathcal C} \approx r_k$, so that cosmological perturbations evolving outside of the horizon are highly {\bf complex}, in this sense \cite{Bhattacharyya:2020rpy,Bhattacharyya:2020kgu,Haque:2021hyw}.

However, despite its high entanglement and complexity,
the squeezed state \eqref{eq:TwoModeSqueezedState} is nonetheless a stabilizer state under Gaussian unitary evolution so that its quantum magic -- the resource required for quantum advantage beyond Clifford evolution -- is exactly zero.
In particular, the operators appearing in $\hat{H}_{\mathbf{k},-\mathbf{k}}$ 
operate within the universal enveloping algebra
of $\mathrm{su}(1,1)$ with the two-mode generators defined as
\begin{equation}
\hat{K}_+ = \hat{c}_{\mathbf{k}}^\dagger \hat{c}_{-\mathbf{k}}^\dagger, \quad \hat{K}_- = \hat{c}_{\mathbf{k}} \hat{c}_{-\mathbf{k}}, \quad \hat{K}_0 = \frac{1}{2} \left( \hat{c}_{\mathbf{k}}^\dagger \hat{c}_{\mathbf{k}} + \hat{c}_{-\mathbf{k}}^\dagger \hat{c}_{-\mathbf{k}} \right).
\end{equation}
As in the single mode case, the rotation and squeezing operators can be written in terms of these generators
\begin{equation}
    \hat {\mathcal S}_{\mathbf k}(r_k,\phi_k) = \exp\left[\frac{r_k}{2} \left(e^{-2i\phi_k}\hat K_- - e^{2i\phi_k}\hat K_+\right)\right]\, , \qquad \hat {\mathcal R}_{\mathbf k}(\theta_k) = \exp\left[-i\theta_k \hat K_0\right]\, .
    \label{eq:SU11SqueezeRotate}
\end{equation}
As in Section \ref{sec:ContinuousStabilizer}, 
we utilize the continuous SU(1,1) group generated by \eqref{eq:SU11SqueezeRotate}
to establish a stabilizer framework. 
Unitary time evolution generated by the two-mode Hamiltonian \eqref{eq:CosmoHamiltonian} 
consists of Gaussian unitaries
(although in practice, unitary evolution will be confined to the SU(1,1) subgroup).
Thus, instead of evolving the infinite-dimensional state vector $|\Psi(\eta)\rangle$, we need only track the evolution of the stabilizer generator.

We take our computational basis to be the two-mode Fock number basis $\{|n_{\mathbf k},n_{-\mathbf k}\rangle\}$. The initial Bunch-Davies vacuum state is stabilized by the two-mode rotation operator
\begin{equation}
    \hat {\mathcal R}_{\mathbf k}(\vartheta) |0_{\mathbf k}, 0_{-\mathbf k}\rangle = |0_{\mathbf k}, 0_{-\mathbf k}\rangle\quad \forall\quad 0<\vartheta\leq 2\pi\, ,
\end{equation}
where the variable $\vartheta$ parameterizes the $U(1)$ stabilizer group, and is different than $\theta_k(\eta)$ which appears in the decomposition of the unitary time evolution \eqref{eq:CosmoUnitary}.
Under unitary evolution $\hat {\mathcal U}_{\mathbf k}$ this generator transforms via conjugation
\begin{equation}
    \hat {\mathcal R}_{\mathbf k}'(\vartheta,\eta) = \hat {\mathcal U}_{\mathbf k}(\eta) \hat {\mathcal R}_{\mathbf k}(\vartheta,\eta) \hat {\mathcal U}_{\mathbf k}^\dagger(\eta) = \exp\left(-i\vartheta \hat {\mathcal K}_0\right)\,,
    \label{eq:TwoModeStabilizer}
\end{equation}
where the transformed stabilizer generator can be expressed in terms of the algebra generators and squeezing parameters
\begin{equation}
\hat{\mathcal{K}}_0(\eta) = \cosh(2r_k) \hat{K}_0 - \sinh(2r_k) \left[ \cos(2\phi_k) \frac{\hat{K}_+ + \hat{K}_-}{2} + i\sin(2\phi_k) \frac{\hat{K}_+ - \hat{K}_-}{2} \right].
\end{equation}
This dynamically updated generator serves as a continuous-variable approach to efficient stabilizer tracking.
Instead of explicitly evolving the infinite-dimensional continuous state vector
we simply track the quantities $r_k(\eta),\phi_k(\eta)$.
This allows us to reconstruct the state $|\Psi(\eta)\rangle_{\mathbf k, -\mathbf k}$ at any time $\eta$ through the exact eigenvalue equation
\begin{equation}
    \hat {\mathcal R}_{\mathbf k}'(\vartheta,\eta)|\Psi(\eta)\rangle_{\mathbf k, -\mathbf k} = \exp\left(-i\vartheta \hat {\mathcal K}_0\right)|\Psi(\eta)\rangle_{\mathbf k, -\mathbf k} = |\Psi(\eta)\rangle_{\mathbf k, -\mathbf k}\, .
\end{equation}

Thus, {\bf the inflationary state is a stabilizer state} at all times at the quadratic level.
Consequently, the evolution of the inflationary perturbations 
remains confined to the Gaussian sector:
the Wigner function is Gaussian, and inflationary perturbations possess zero quantum magic.
This is consistent with describing the time-dependent two-mode squeezed state \eqref{eq:TwoModeSqueezedState} entirely in terms of the three time-dependent functions $r_k(\eta),\phi_k(\eta),\theta_k(\eta)$ for each mode pair $(\mathbf k,-\mathbf k)$,
despite the large squeezing and entanglement generated during inflation. 
A breakdown of this stabilizer approach towards inflationary perturbations would 
require interactions that generate an uncontrolled hierarchy of higher-order correlators and cumulants.
Nonetheless, re-framing the time-evolution of the quantum state within the stabilizer state formalism allows us to directly 
connect inflationary dynamics to
the continuous-variable version of the Gottesman-Knill theorem, 
and provides a natural framework for analyzing the effects of primordial non-Gaussianity.

In order to construct a well-defined Wigner probability density, we place the system in a comoving box of finite volume $V$, transitioning from continuous Fourier modes to discrete variables $\hat v_{\mathbf k} \rightarrow \hat v_{\mathbf k}/\sqrt{V}$ and $\hat p_{\mathbf k} \rightarrow \hat p_{\mathbf k}/\sqrt{V}$.
We decompose the Mukhanov variable and its conjugate momentum into ``real" and ``imaginary" Hermitian variables $\hat q_{\mathbf k}^{R,I}, \hat p_{\mathbf k}^{R,I}$
\begin{align}
    \hat v_{\mathbf k} &= \frac{\hat q_{\mathbf k}^R + i \hat q_{\mathbf k}^I}{\sqrt{2}}, \hspace{.2in} \hat p_{\mathbf k} = \frac{\hat p_{\mathbf k}^R + i\hat p_{\mathbf k}^I}{\sqrt{2}}\, .
    \label{eq:RealImqp}
\end{align}
The reality condition $\hat v_k^\dagger = \hat v_{-k}$ implies that  the ``real" and ``imaginary" parts are even/odd under parity transformations $(\hat q_k^R)^\dagger = \hat q_{-k}^R =\hat q_k^R$ and $(\hat q_k^I)^\dagger = -\hat q_{-k}^I = \hat q_k^I$ (with similar relations for $\hat p_K^{R,I}$).
Due to the isotropy of the background, 
the real and imaginary sectors decouple $\hat {\mathcal H}_{\mathbf k,-\mathbf k} = \hat {\mathcal H}_{\mathbf k}^R + \hat {\mathcal H}_{\mathbf k}^I$, and the covariance matrix of each sector becomes (in terms of the dimensionless variables $\tilde q_{\mathbf k}=\sqrt k\,q_{\mathbf k}$, $\tilde p_{\mathbf k}=p_{\mathbf k}/\sqrt k$)
\begin{equation}
    \tilde{\mathbf\Sigma}_{R,I}=\frac12\begin{pmatrix}
    \cosh2r_k-\cos2\phi_k\sinh2r_k & \sin2\phi_k\sinh2r_k\\
    \sin2\phi_k\sinh2r_k & \cosh2r_k+\cos2\phi_k\sinh2r_k\end{pmatrix},
\end{equation}
with $\det\tilde{\mathbf\Sigma}_{R,I}=\tfrac14$ and principal-axis variances \eqref{eq:PrincipalVariances} $\lambda_\pm=\tfrac12e^{\pm2r_k}$.
Rotating each mode to the principal axes $(X_{\mathbf k},Y_{\mathbf k})$ following \eqref{eq:PrincipalAxes}, the Gaussian Wigner function factorizes,
\begin{equation}
    W_{\rm Gauss}=\prod_{\mathbf k}\,\prod_{j=R,I}\frac{1}{\pi}
    \exp\left[-\frac{(X^j_{\mathbf k})^2}{e^{2r_k}}-\frac{(Y^j_{\mathbf k})^2}{e^{-2r_k}}\right].
    \label{eq:InflationWigner}
\end{equation}
On superhorizon scales $\phi_k\to-\pi/2+k|\eta|$, so that the major axis is aligned with the field, $X_{\mathbf k}\simeq-\tilde q_{\mathbf k}$, up to corrections of order $k|\eta| \ll 1$.

In the next section, we will use the cumulant expansion developed in Section \ref{sec:Cumulant} to go beyond the Gaussian level, and probe the corresponding amount of non-stabilizerness through magic monotones.

\subsection{Non-Gaussianity and Magic}
\label{sec:NonGaussMagic}

Inflationary perturbations at the quadratic level are Gaussian, and as such are stabilizer states with vanishing quantum magic.
In order to investigate the properties of quantum magic during inflation, we will thus need to go beyond the quadratic level and include self-interactions of the curvature perturbation.
Cubic interaction terms of the curvature perturbation $\zeta$ take the form \cite{Maldacena:2002vr,Babich:2004gb,Chen:2006nt,Adshead:2011bw}
\begin{align}
    S_3 = M_{pl}^2 \int dt d^3x \Big[&a^3 \epsilon^2 \zeta\dot \zeta^2 + a \epsilon^2 \zeta (\partial \zeta)^2 - 2 a\epsilon \dot \zeta (\partial \zeta)(\partial \chi) + a^3 \epsilon (\dot\epsilon - \dot\eta)\zeta^2 \dot \zeta \nonumber \\
    & + \frac{\epsilon}{2a} \partial \zeta \partial \chi \partial^2 \chi +\frac{d}{dt} \left(a^3\epsilon(\epsilon-\eta) \dot \zeta \zeta^2\right)\Big]\,,
    \label{eq:cubicAction}
\end{align}
where $\chi = a^2 \epsilon \partial^{-2} \dot \zeta$, $\epsilon,\eta$ are the slow roll parameters, and spatial derivatives are contracted as $\partial^2 = \delta^{ij} \partial_i \partial_j$.
Different terms in the cubic action \eqref{eq:cubicAction} give different ``shapes'' of the momentum-space three-point function for the amplitude $\langle \zeta_{\mathbf{k}_1}\zeta_{\mathbf{k}_2}\zeta_{\mathbf{k}_3}\rangle$, with slightly different coefficients \cite{Babich:2004gb,Chen:2006nt}.
As a simplified ansatz, three-point functions of the ``local'' type can be generated by a curvature perturbation consisting of a non-linear correction to the Gaussian field \cite{Maldacena:2002vr}
\begin{equation}
    \zeta = \zeta_g + \frac{3}{5} f_{NL} \left(\zeta_g^2-\langle \zeta_g^2\rangle\right)\,.
    \label{eq:fNL}
\end{equation}
The non-linearity parameter $f_{NL}$ controls the strength of the non-Gaussianity; in slow-roll inflation models, $f_{NL} \sim {\mathcal O}(\epsilon,\eta) \ll 1$, while observational limits place  upper bounds on the amplitude of non-Gaussianity $f_{NL} \lesssim {\mathcal O}(1-10)$ \cite{Planck:2019kim}.
The non-linear term \eqref{eq:fNL} is a small correction to the two-point function when $f_{NL} \langle \zeta_g\rangle^{1/2} \ll 1$.
The three-point function corresponding to \eqref{eq:fNL} is commonly expressed in terms of the non-linearity parameter $f_{NL}$ \cite{Maldacena:2002vr,Babich:2004gb,Chen:2006nt}
\begin{equation}
    \langle \zeta_{\mathbf{k}_1}\zeta_{\mathbf{k}_2}\zeta_{\mathbf{k}_3}\rangle =
    \frac{6}{5} f_{NL} (2\pi^2)^2 \left(\Delta^2_\zeta(k)\right)^2 \left(\frac{\sum_i k_i^3}{\prod_i k_i^3}\right)\ (2\pi)^3 \delta^3\left(\mathbf{k}_1+\mathbf{k}_2+\mathbf{k}_3\right)\,,
\end{equation}
where $\Delta^2_\zeta(k)$ is the dimensionless power spectrum of curvature perturbations constant on large scales, defined in \eqref{eq:SqueezedTwoPointCurvature}.
More generally, the three point function can be written in terms of an \emph{effective} non-linearity parameter
\begin{equation}
    \langle \zeta_{\mathbf{k}_1}\zeta_{\mathbf{k}_2}\zeta_{\mathbf{k}_3}\rangle  = f_{NL}^{\rm eff} \left(\Delta^2_\zeta(k)\right)^2 F(\mathbf k_1,\mathbf k_2,\mathbf k_3) \ (2\pi)^3 \delta^3\left(\mathbf{k}_1+\mathbf{k}_2+\mathbf{k}_3\right)\, ,
    \label{eq:fNLeff}
\end{equation}
where $F(\mathbf k_1,\mathbf k_2,\mathbf k_3)$ is a function of the momenta of degree $-6$.
Expressed in terms of the real and imaginary Mukhanov variables \eqref{eq:RealImqp}, the three-point function in terms of the non-linearity parameter \eqref{eq:fNLeff} becomes 
\begin{equation}
    \langle q_{\mathbf k_1}^{(R,I)}  q_{\mathbf k_2}^{(R,I)} q_{\mathbf k_3}^{(R,I)}\rangle  = f_{NL}^{\rm eff}\left(\Delta^2_\zeta(k)\right)^2 (M_p a(\eta) \sqrt{2\epsilon})^3\ F(\mathbf k_1,\mathbf k_2,\mathbf k_3)\ (2\pi)^3 \delta^3\left(\mathbf{k}_1+\mathbf{k}_2+\mathbf{k}_3\right) \, .
    \label{eq:qThreePoint}
\end{equation}
Because of the parity of the $q_{\mathbf k}^{(R,I)}$, three-point functions with an odd number of
$q^I$ factors vanish; those with zero or two, such as $\langle q^Rq^Rq^R\rangle$ and
$\langle q^Rq^Iq^I\rangle$, are both non-zero and of equal magnitude.

We note that observations do not measure the quantum expectation values $\langle\hat\zeta_{\mathbf k'}\hat\zeta_{\mathbf k}\rangle$ and $\langle\hat\zeta_{\mathbf k_1}\hat\zeta_{\mathbf k_2}\hat\zeta_{\mathbf k_3}\rangle$ directly. 
Instead, the CMB provides a single realization of the temperature anisotropy, from which we construct estimators of the angular power spectrum and bispectrum \cite{Planck:2019kim}.
Connecting these to primordial correlators requires standard assumptions of statistical homogeneity and isotropy, as well as a linear transformation from $\zeta$ to the anisotropy coefficients and the identification of our sky with a typical realization of the underlying ensemble.
We will take the observed sky to be a particular realization drawn from the ensemble described by the pure state $|\Psi\rangle$, but do not address the corresponding measurement problem in cosmology \cite{Perez:2005gh,Kiefer:2008ku,Martin:2015qta,Sudarsky:2009za}.
Decoherence (see e.g.~\cite{Zurek_Coherent_States_Decoherence,Campo:2004sz,Martineau:2006ki,Burgess:2006jn,Kiefer:2008ku,Burgess:2014eoa,Nelson:2016kjm,Burgess:2022nwu,Bhattacharyya:2024duw,Bhattacharyya:2025cxv,Micheli:2025yux}, among others) produces a mixed state, for which Hudson's Theorem does not apply, and it would be interesting to study in future work how Wigner negativity is affected by different decoherence models.
Since the observed power spectrum and bispectrum only constrain the statistics of the field, and do not measure the quantum expectation values directly, we will use observational bounds on $\Delta_\zeta$ and $f_{NL}$, interpreted within these standard assumptions, to infer the Wigner negativity of the quantum inflationary  perturbations.

The bispectrum \eqref{eq:fNLeff},\eqref{eq:qThreePoint} couples three distinct Fourier modes.
To apply the single-mode analysis of Section \ref{sec:Cumulant}, we first rotate to the principal axis variables for each Fourier mode $(X_{\mathbf k},Y_{\mathbf k})$ following \eqref{eq:PrincipalAxes}.
For simplicity, we consider modes forming an equilateral triangle
$\mathbf k_1+\mathbf k_2+\mathbf k_3=0$ with $|\mathbf k_i|=k$ so that the three modes share the same squeezing parameters $r_k, \phi_k$ and Gaussian Wigner function \eqref{eq:InflationWigner}.
In the superhorizon limit $r_k \gg 1$, $\phi_k \simeq -\pi/2$ the third-order cumulant along the principal axis variables $X_i$ reduces to the bispectrum
\begin{align}
C_{X_1X_2X_3} &= \sin^3\phi_k\,C_{\tilde q_1\tilde q_2\tilde q_3} +3\sin^2\phi_k\cos\phi_k\,
C_{\tilde q_1\tilde q_2\tilde p_3} +3\sin\phi_k\cos^2\phi_k\,C_{\tilde q_1\tilde p_2\tilde p_3} +\cos^3\phi_k\,C_{\tilde p_1\tilde p_2\tilde p_3} \\
&\approx -k^{3/2} C_{q_1q_2q_3}\, . \label{eq:XCumulant}
\end{align}
By focusing on the collective coordinates
\begin{equation}
    \mathcal X=\frac{X_1+X_2+X_3}{\sqrt3}\, .\qquad \mathcal Y=\frac{Y_1+Y_2+Y_3}{\sqrt3}\, .
\end{equation}
we can model the non-Gaussianity as a single squeezed mode\footnote{There are two other collective modes created from orthogonal combinations of the $X_i$. However, the largest third-order cumulant is in the homogeneous combination $\mathcal X$, and the contribution of these additional modes will simply contribute an ${\mathcal O}(1)$ numerical factor to the negativity with the same parametric dependence on $f_{NL},\Delta_\zeta$. Since we find the negativity to be highly suppressed, we will not be concerned with these numerical coefficients.} as in Section \ref{sec:Cumulant} with a dimensionless skewness \eqref{eq:GammaDef}
\begin{equation}
    \gamma_{\mathcal X}=\frac{C_{\mathcal X\mathcal X\mathcal X}}{\lambda_+^{3/2}} = \frac{2}{\sqrt3}\frac{C_{X_1X_2X_3}}{\lambda_+^{3/2}}
    \simeq\frac{2}{\sqrt3}\frac{\langle q_1^Rq_2^Rq_3^R\rangle}{\sigma_q^3}
    =\frac{2}{\sqrt3}\frac{\langle\zeta^R_1\zeta^R_2\zeta^R_3\rangle}{\langle\zeta^2\rangle^{3/2}}\, ,
    \label{eq:GammaCollective}
\end{equation}
so that the dimensionless skewness $\gamma_{\mathcal X}$ is proportional to the bispectrum.
Since the factors of $z=aM_p\sqrt{2\epsilon}$ cancel, $\gamma_{\mathcal X}$ is independent of time on superhorizon scales.
Inserting our expressions for the power spectrum and bispectrum, we find
\begin{equation}
    \gamma_{\mathcal X}=\mathcal A\,f^{\rm eff}_{NL}\Delta_\zeta\, ,\qquad
    \mathcal A=\sqrt{\frac23}\,\frac{k^6F(k,k,k)}{(2\pi^2)^{3/2}}\,\frac{1}{\sqrt{k^3V}}\, .
    \label{eq:GammaMatching}
\end{equation}
For the smallest box volume containing the modes $V=(2\pi/k)^3$ and local non-Gaussianity at the equilateral point, $k^6F=\tfrac{18}{5}(2\pi^2)^2$, we have $\mathcal A\simeq0.83$. More generally, $\mathcal A \sim {\mathcal O}(1)$ so that the dimensionless skewness $\gamma_{\mathcal X}$ is parametrically controlled by the non-Gaussianity parameter $f_{NL}^{\rm eff}$ and the power spectrum.

Under this quasi-single mode analysis, with $C_{\mathcal X\mathcal X\mathcal X}$ as the only higher-order cumulant, the characteristic function of the collective mode parallels that of \eqref{eq:TruncatedGeneratingFunction}
\begin{equation}
    \chi(\xi_{\mathcal X},\xi_{\mathcal Y})=\exp\left[-\frac12\lambda_+\xi_{\mathcal X}^2-\frac12\lambda_-\xi_{\mathcal Y}^2-\frac{i}{6}C_{\mathcal X\mathcal X\mathcal X}\,\xi_{\mathcal X}^3\right].
    \label{eq:TruncatedGeneratingFunctionInflation}
\end{equation}
and the non-Gaussian Wigner function takes the form \eqref{eq:WignerAiryForm}
\begin{equation}
    W(x,y)=\frac{e^{-y^2/2}}{\sqrt{2\pi}}\,{\mathcal A}_{\gamma_{\mathcal X}}(x)\times(\text{spectators})\, ,
    \label{eq:CollectiveAiry}
\end{equation}
in terms of the scaled variables $x = {\mathcal X}/\sqrt{\lambda_+},\ y={\mathcal Y}/\sqrt{\lambda_-}$, where the scaled Airy function is \eqref{eq:ScaledAiryFunction}, and the spectators include the other collective modes that we have neglected..
The negativity of the collective mode thus reduces to the same integral \eqref{eq:NegativityAiryWigner} over the scaled Airy function
\begin{align}
    \mathcal N & =-2\int d\mathcal X\,d\mathcal Y\,[W]_-
    =-2\int dx\,\big[{\mathcal A}_{\gamma_{\mathcal X}}(x)\big]_- \\
    & \approx  2^{5/3} \gamma_{\mathcal X}^{4/3}\ |{\rm Ai}'(a_1)|\ \exp\left[-\frac{1}{6\gamma_{\mathcal X}^2} - \frac{|a_1|}{2^{1/3} \gamma_{\mathcal X}^{2/3}}\right] \left[1-2^{2/3} |a_1| \gamma_{\mathcal X}^{4/3} + {\mathcal O}{(\gamma_{\mathcal X}^2})\right]\, .
    \label{eq:NegativityCollective}
\end{align}
With $\Delta_\zeta^2\simeq 2.1\times10^{-9}$ and $\mathcal A \simeq0.83$, we have $\gamma_{\mathcal X}\simeq 4\times 10^{-4}$ for the edge of observationally allowed non-Gaussianity $f^{\rm eff}_{NL}=10$ \cite{Planck:2019kim}, and $\gamma_{\mathcal X}\simeq 4\times 10^{-7}$ for slow-roll inflation with $f^{\rm eff}_{NL}\sim 10^{-2}$.
As a result, even at the observationally allowed bound the exponential suppression in \eqref{eq:NegativityCollective} is $\sim e^{-10^{6}} \ll 1$ and Wigner negativity is highly suppressed; the suppression is even more extreme for slow roll inflation.

It is useful to consider how generic the above result is across models of inflation.
We found that a primarily Gaussian distribution of cosmological perturbations with bispectrum of the form \eqref{eq:fNLeff} should have a strongly suppressed Wigner negativity of the form \eqref{eq:NegativityCollective}.
For example, while inflationary models with non-canonical kinetic terms (such as $k$-inflation \cite{Armendariz-Picon:1999hyi,Chen:2006nt} or DBI-Inflation \cite{Alishahiha:2004eh}) produce enhanced non-Gaussianities, observational constraints on $f_{NL}$ and suppression by the power spectrum still lead to a small dimensionless skewness parameter $\gamma_{\mathcal X} \ll 1$ and suppressed negativity.
However, a non-linear transformation between the canonically normalized and physical field variables, such as occurs during ultra-slow-roll inflation, may generate Wigner negativity by warping the otherwise Gaussian Wigner function \cite{Ireland:2026txt}.
Ultra-slow-roll dynamics of this kind are also characteristic of models that produce primordial black holes, where the curvature perturbation can develop large, non-perturbative non-Gaussianity \cite{Atal:2018neu,Franciolini:2018vbk,Cai:2022erk}.
Alternatively, multifield inflationary models \cite{Byrnes:2010em} may generate Wigner negativity through the transfer of isocurvature perturbations into the curvature perturbation.
Finally, interactions of the cosmological perturbations with environmental degrees of freedom force the system into a mixed state.
While decoherence is generally expected to preserve the Gaussianity of the state, whether specific interactions can induce significant negativity is an interesting avenue for future study.

\section{Discussion}
\label{sec:Discussion}

In this paper, we have examined the existence of stabilizers and quantum magic in the quantum cosmological perturbations generated during inflation. 
By generalizing the continuous-variable stabilizer formalism to a framework suited for cosmology, we demonstrated that the highly squeezed vacuum states produced during inflation are stabilizer states.
As a result, the quadratic dynamics of inflation remain confined to the stabilizer sector, with vanishing Wigner negativity and zero quantum magic.
This is consistent with the fact that these states are fully characterized by the squeezing and rotation parameters $r_k(\eta), \phi_k(\eta)$, and $\theta_k(\eta)$,
despite exhibiting large squeezing, entanglement, and circuit complexity.
Pure quantum states with primordial non-Gaussianities, as measured by the non-Gaussianity parameter $f_{NL}$ of the bispectrum, introduce Airy oscillations in the corresponding Wigner function controlled by the dimensionless skewness parameter $\gamma_{\mathcal X} = {\mathcal A} f_{NL} \Delta_\zeta$, with $\Delta_\zeta$ the curvature power spectrum. 
The resulting Wigner negativity is highly suppressed ${\mathcal N} \sim \gamma_{\mathcal X}^{4/3} \exp[-1/(6\gamma_{\mathcal X}^2)]$, so that even for the largest non-Gaussianities consistent with observations $f_{NL} \sim {\mathcal O}(10)$ the negativity is vanishingly small.

It is instructive to compare the small quantum magic of cosmological perturbations with quantum–information–theoretic measures such as 
quantum circuit complexity\footnote{It is also interesting to consider other quantum–information–theoretic quantities applied to cosmological perturbations, such as quantum discord \cite{Martin:2015qta} and Bell-inequality violations \cite{Martin:2017zxs,Martin:2019wta,Martin:2022kph}.}.
While we have shown that Wigner negativity is exponentially suppressed, circuit complexity -- a measure of the number of quantum gates necessary to construct the resulting unitary evolution -- behaves oppositely.
Previous work \cite{Bhattacharyya:2020rpy,Bhattacharyya:2020kgu,Haque:2021hyw} has shown that the circuit complexity of cosmological perturbations grows linearly with the squeezing parameter, ${\mathcal C}\approx r_k$, and thus becomes large even though the Wigner negativity and quantum magic vanishes.

It is interesting that the entangled quantum perturbations produced during inflation exhibit growth of circuit complexity while simultaneously possessing vanishingly small quantum magic, and that these properties appear to map to the physical requirements for successful inflation.
Accelerated expansion requires an approximately constant Hubble expansion rate with $\epsilon = -\dot H/H^2 \ll 1$.
As shown in \cite{Bhattacharyya:2020kgu}, quasi–de Sitter evolution with small $\epsilon$ yields the maximal allowed rate of complexity growth among backgrounds consistent with the null energy condition.
In parallel, our analysis shows that Wigner negativity is highly sensitive to the inflaton’s self-interactions. 
Prolonged inflation with $\sim 50-60$ e-folds requires a rolling homogeneous scalar field on a nearly flat potential with $|\eta| = |\dot \epsilon/(\epsilon H)| \ll 1$. 
From this perspective, it is interesting to view the cosmological conditions for sustained accelerated expansion as information-theoretic conditions of large complexity growth and vanishing magic.

This pairing suggests a deeper connection between these information-theoretic measures and the notion of ``classicality" in cosmology. 
The quantum states most often associated with classical behavior -- coherent states and squeezed vacuum states -- are Gaussian. 
They are naturally characterized by high circuit complexity (due to squeezing or displacement) but zero quantum magic (due to Gaussianity). 
This interplay also points toward applications to decoherence and the quantum-to-classical transition. 
Although interactions with environmental degrees of freedom are necessary to decohere super-Hubble modes, observational constraints \cite{Planck:2019kim} suggest that these interactions preserve the near-Gaussian character of the perturbations. 
In that case, decoherence likely sustains or further suppressed the already small magic of the system.
It would be valuable to determine how specific decoherence mechanisms (see e.g.~\cite{Zurek_Coherent_States_Decoherence,Campo:2004sz,Martineau:2006ki,Burgess:2006jn,Kiefer:2008ku,Burgess:2014eoa,Nelson:2016kjm,Burgess:2022nwu,Bhattacharyya:2024duw,Bhattacharyya:2025cxv,Micheli:2025yux}, among others) affect the relationship between these quantities, such as whether interactions can slow the growth of complexity while maintaining the suppression of Wigner negativity.
Validating the relationships between these quantities may clarify whether minimal magic and high complexity growth serve as robust markers of the emerging classical behavior.

\section*{Acknowledgments}
We would like to thank Rob Myers for discussions.
SSH is supported in part by the National Institute for Theoretical and Computational Sciences
of South Africa (NITheCS). BU would like to thank NITheCS for funding for travel to the University of Cape Town where part of this work was completed.

\appendix

\section{Cumulant Expansion Details}
\label{app:Cumulant}

In the main text, we outlined the calculation of corrections to the Gaussian Wigner function from third-order and fourth-order cumulants in terms of the multivariate Hermite polynomials \eqref{eq:MultivariateHermite}. In this appendix we write out these terms explicitly.

\subsection{Third-order Cumulant}

The linear third-order cumulant correction to the Wigner function takes the form of a cubic polynomial times the Gaussian Wigner function
\begin{equation}
\Delta W^{(3)}(\vec z)
= W_G(\vec{z}) \frac{1}{3!} C_{abc} H_{abc}(\vec z; \mathbf{\Sigma})
= W_G(\vec z)\Big(-\tfrac{1}{2}\,V_\gamma\,z^\gamma
+\tfrac{1}{6}\,\kappa_{\alpha\beta\gamma}\,z^\alpha z^\beta z^\gamma\Big)\,,
\end{equation}
where the coefficients are
\begin{equation}
V_\gamma \equiv C_{ijk}A_{ij}A_{k\gamma},\qquad
\kappa_{\alpha\beta\gamma}\equiv C_{ijk}A_{i\alpha}A_{j\beta}A_{k\gamma}.
\label{eq:HermiteCoefficients}
\end{equation}
and $\mathbf{A} = \mathbf{\Sigma}^{-1}$.
The third-order correction $\Delta W^{(3)}(\vec z)$ skews the Gaussian distribution.  
For a single mode ($N=1$), the correction is determined by a cubic polynomial in $q, p$
\begin{equation}
\Delta W^{(3)}(q,p)
= W_G(q,p)\Big[-\tfrac{1}{2}(v_q\,q+v_p\,p) + \tfrac{1}{6}\big(\kappa_{Q^3}q^3+\kappa_{Q^2P}q^2p+\kappa_{QP^2}q p^2+\kappa_{P^3}p^3\big)\Big]\, ,
\end{equation}
whose coefficients are
\begin{align}
v_q &= \frac{1}{\Delta^2}\Big[\,\sigma_{p}^4\,C_{qqq}
-3\,\sigma_{p}^2\sigma_{qp}\,C_{qqp}
+\big(2\sigma_{qp}^2+\sigma_{p}^2\sigma_{q}^2\big)\,C_{qpp}
-\sigma_{q}^2\sigma_{qp}\,C_{ppp}\Big];\\[6pt]
v_p &= \frac{1}{\Delta^2}\Big[\,\sigma_{q}^2\,C_{ppp}
-3\,\sigma_{q}^2\sigma_{qp}\,C_{qpp}
+\big(2\sigma_{qp}^2+\sigma_{p}^2\sigma_{q}^2\big)\,C_{qqp}
-\sigma_{p}^2\sigma_{qp}\,C_{qqq}\Big],
\end{align}
and
\begin{align}
\kappa_{Q^3} &= \frac{1}{\Delta^3}\Big[\,\sigma_{p}^6\,C_{qqq}
-3\sigma_{p}^4\sigma_{qp}\,C_{qqp}
+3\sigma_{p}^2\sigma_{qp}^2\,C_{qpp}
-\sigma_{qp}^3\,C_{ppp}\Big];\\[6pt]
\kappa_{P^3} &= \frac{1}{\Delta^3}\Big[\,\sigma_{q}^6\,C_{ppp}
-3\sigma_{q}^4\sigma_{qp}\,C_{qpp}
+3\sigma_{q}^2\sigma_{qp}^2\,C_{qqp}
-\sigma_{qp}^3\,C_{qqq}\Big];\\[8pt]
\kappa_{Q^2P} &= \frac{3}{\Delta^3}\Big[
-\,\sigma_{p}^4\sigma_{qp}\,C_{qqq}
+(\sigma_{p}^4\sigma_{q}^2+2\sigma_{p}^2\sigma_{qp}^2)\,C_{qqp}\\
&\qquad\qquad\qquad
-(2\sigma_{p}^2\sigma_{qp}\sigma_{q}^2+\sigma_{qp}^3)\,C_{qpp}
+\sigma_{qp}^2\sigma_{q}^2\,C_{ppp}
\Big];\\[10pt]
\kappa_{QP^2} &= \frac{3}{\Delta^3}\Big[
\sigma_{qp}^2\sigma_{p}^2\,C_{qqq}
-(2\sigma_{q}^2\sigma_{qp}\sigma_{p}^2+\sigma_{qp}^3)\,C_{qqp}\\
&\qquad\qquad\qquad
+(\sigma_{q}^4\sigma_{p}^2+2\sigma_{q}^2\sigma_{qp}^2)\,C_{qpp}
-\sigma_{q}^4\sigma_{qp}\,C_{ppp}
\Big].
\end{align}
with $\Delta=\sigma_{q}^2\sigma_{p}^2-\sigma_{qp}^2 = 1/4$ for pure squeezed vacuum states.

\subsection*{Fourth-order Cumulant}

The linear fourth-order cumulant contribution to the Wigner function from \eqref{eq:WignerCorrected} can be written as a polynomial in the phase space variables times the Gaussian Wigner function
\begin{align}
\Delta W^{(4)}(\vec z)
&= W_G(\vec{z})\frac{1}{4!} C_{abcd} H_{abcd}(\vec z, \mathbf{\Sigma}) \nonumber \\
& =W_G(y)\left[
\underbrace{\frac{1}{8}C_{abcd}A_{ab}A_{cd}}_{\text{constant}}
-
\underbrace{\frac{1}{4}C_{abcd}A_{ab}A_{c\alpha}A_{d\beta}z^\alpha z^\beta}_{\text{quadratic in }z}
+
\underbrace{\frac{1}{24}C_{abcd}A_{a\alpha}A_{b\beta}A_{c\gamma}A_{d\delta}z^\alpha z^\beta a^\gamma z^\delta}_{\text{quartic in }y}
\right].
\label{eq:Wigner4thOrder}
\end{align}
where as before repeated indices are summed, $\mathbf A = \mathbf{\Sigma}^{-1}$, and we used the definition of the multivariate Hermite polynomials \eqref{eq:MultivariateHermite}.

In the case of a single mode $N=1$, the fourth-order correction takes the form
\begin{align}
    \Delta W^{(4)}(q,p) = W_G(q,p)\Big[ &C_0
+\tfrac12\big(M_{qq}q^2 + 2M_{qp}qp + M_{pp}p^2\big)\nonumber \\
&+\tfrac{1}{24}\big(D_{q^4}q^4 + D_{q^3p}q^3p + D_{q^2p^2}q^2p^2 + D_{qp^3}qp^3 + D_{p^4}p^4\big)
\Big]\,,
\label{eq:Wigner4thOrder1}
\end{align}
where
\begin{align}
C_0 &= \frac{1}{8\Delta^{2}}\Big(
        C_{pppp}\sigma_{q}^{4}
        -4C_{qppp}\sigma_{qp}\sigma_{q}^2
        +6C_{qqpp}\sigma_{p}^2\sigma_{q}^2
        -4C_{qqqp}\sigma_{p}^2\sigma_{qp}
        +C_{qqqq}\sigma_{pp}^{4}
    \Big)\,; \\
M_{qq} &= \frac{-1}{4\Delta^{3}}\Big(
        C_{qqqq}\sigma_{p}^{6}
        -3C_{qqqp}\sigma_{p}^{4}\sigma_{qp}
        +3C_{qqpp}\sigma_{p}^2\sigma_{qp}^{2}
        -C_{qppp}\sigma_{qp}^{3}
    \Big)\, ; \\
M_{qp} &= \frac{1}{4\Delta^{3}}\Big(
        C_{qqqq}\sigma_{p}^{4}\sigma_{qp}
        -2C_{qqqp}\sigma_{p}^2\sigma_{qp}^{2}
        +C_{qqpp}(\sigma_{p}^2\sigma_{qq}\sigma_{qp}+\sigma_{qp}^{3})
        -2C_{qppp}\sigma_{qp}^{2}\sigma_{q}^2
        +C_{pppp}\sigma_{qp}\sigma_{qq}^{4}
    \Big)\,; \\
M_{pp} &= \frac{1}{4\Delta^{3}}\Big(
        C_{qqqq}\sigma_{qp}^{3}
        -3C_{qqqp}\sigma_{qp}^{2}\sigma_{q}^2
        +3C_{qqpp}\sigma_{qp}\sigma_{qq}^{4}
        -C_{qppp}\sigma_{qq}^{6}
    \Big)\,; \\
D_{q^{4}} &= \frac{1}{24\Delta^{4}}\Big(
        C_{qqqq}\sigma_{pp}^{4}
        -4C_{qqqp}\sigma_{p}^{6}\sigma_{qp}
        +6C_{qqpp}\sigma_{p}^{6}\sigma_{qp}^{2}
        -4C_{qppp}\sigma_{p}^2\sigma_{qp}^{3}
        +C_{pppp},\sigma_{qp}^{4}
    \Big)\,; \\
D_{p^{4}} &= \frac{1}{24\Delta^{4}}\Big(
        C_{qqqq}\sigma_{qp}^{4}
        -4C_{qqqp}\sigma_{qp}^{3}\sigma_{q}^2
        +6C_{qqpp}\sigma_{qp}^{2}\sigma_{q}^{4}
        -4C_{qppp}\sigma_{qp}\sigma_{q}^{6}
        +C_{pppp},\sigma_{q}^{8}
    \Big)\,; \\
D_{q^{3}p} &= \frac{1}{24\Delta^{4}}\Big(
        C_{qqqq}\sigma_{p}^{6}\sigma_{qp}
        -3C_{qqqp}\sigma_{p}^{3}\sigma_{qp}^{2}
        +3C_{qqpp}\sigma_{p}^2\sigma_{qp}^{3}
        -C_{qppp}\sigma_{qp}^{4}\\
    &\qquad\qquad
        +C_{qqqp}\sigma_{p}^{8}
        -3C_{qqpp}\sigma_{p}^{6}\sigma_{qp}
        +3C_{qppp}\sigma_{p}^{4}\sigma_{qp}^{2}
        -C_{pppp}\sigma_{p}^2\sigma_{qp}^{3}
    \Big)\,; \\
D_{q^{2}p^{2}} &= \frac{1}{24\Delta^{4}}\Big(
        C_{qqqq}\sigma_{p}^{4}\sigma_{qp}^{2}
        -2C_{qqqp}\sigma_{p}^2\sigma_{qp}^{3}
        +C_{qqpp}( \sigma_{p}^2\sigma_{q}^2\sigma_{qp}^{2} + 2\sigma_{p}^{4}\sigma_{q}^2\sigma_{qp} )\\
    &\qquad\qquad
        -2C_{qppp}\sigma_{p^2}\sigma_{q}^2\sigma_{qp}^{2}
        +C_{pppp}\sigma_{q}^{4}\sigma_{p}^2
    \Big)\,; \\
D_{qp^{3}} &= \frac{1}{24\Delta^{4}}\Big(
        C_{qqqq}\sigma_{p^2}\sigma_{qp}^{3}
        -3C_{qqqp}\sigma_{p^2}\sigma_{qp}^{2}\sigma_{q}^2
        +3C_{qqpp}\sigma_{p}^2\sigma_{qp}\sigma_{q}^{4}
        -C_{qppp}\sigma_{q}^{6}\sigma_{qp}\\
    &\qquad\qquad
        +C_{qqqp}\sigma_{p}^{4}\sigma_{q}^2\sigma_{qp}
        -3C_{qqpp}\sigma_{p}^{4}\sigma_{q}^{4}
        +3C_{qppp}\sigma_{p}\sigma_{q}^{6}
        -C_{pppp}\sigma_{q}^{8}
    \Big)\,.
\end{align}

In the case where only $C_{qqqq}$ is non-zero (or alternatively, where $C_{qqqq}$ dominates over the other terms), the correction \eqref{eq:Wigner4thOrder1} becomes
\begin{align}
    C_0 &= \frac{C_{qqqq}\sigma_{p}^{4}}{8\Delta^{2}}\,; \\
    M_{qq} &= -\frac{C_{qqqq}\sigma_{p}^{6}}{4\Delta^{3}},\qquad
    M_{qp} = \frac{C_{qqqq}\sigma_{p}^{4}\sigma_{qp}}{4\Delta^{3}},\qquad
    M_{pp} = \frac{C_{qqqq}\sigma_{p}^2\sigma_{qp}^{2}}{4\Delta^{3}} \,; \\
    D_{q^{4}} &= \frac{C_{qqqq}\sigma_{p}^{8}}{24\Delta^{4}},\qquad
    D_{q^{3}p} = \frac{C_{qqqq}\sigma_{p}^{6}\sigma_{qp}}{24\Delta^{4}},\qquad
    D_{q^{2}p^{2}} = \frac{C_{qqqq}\sigma_{p}^{4}\sigma_{qp}^{2}}{24\Delta^{4}};\\
    D_{qp^{3}} &= \frac{C_{qqqq}\sigma_{p}^2\sigma_{qp}^{3}}{24\Delta^{4}},\qquad
    D_{p^{4}} = \frac{C_{qqqq}\sigma_{qp}^{4}}{24\Delta^{4}}.
\end{align}
Transforming to the variable basis
\begin{align}
    Q = q, \qquad P = p - \frac{\sigma_{qp}}{\sigma_q^2} q\, .
    \label{eq:SymplecticQP}
\end{align}
and taking only $C_{QQQQ} = C_{qqqq}$ to be non-zero, the corresponding fourth-order cumulant correction to the Wigner function is
\begin{equation}
W(Q,P) = W_G(Q,P) \left[1 + \frac{C_{qqqq}}{\sigma_q^4} \left(\frac{Q^4}{\sigma_q^4} - 6 \frac{Q^2}{\sigma_q^2} + 3\right)\right]\, ,
\label{eq:4thOrderWignerPoly}
\end{equation}
where as before, $\sigma_q^2 \sim e^{2r}\gg 1$.
It is straightforward to see that \eqref{eq:4thOrderWignerPoly} will not be negative for 
$C_{qqqq}/\sigma_q^4 \ll 1$;
therefore, there is no contribution to the negativity from this correction to the Wigner function.

\section{Wigner Function of a Cubic Phase Gate State}
\label{app:CubicGate}
In this appendix, we construct a state from a cubic phase gate acting on a Gaussian and show that its Wigner function has a similar Airy function structure, and Wigner negativity, as the Wigner function \eqref{eq:WignerAiryForm} constructed from keeping only the third-order cumulant $C_{XXX}$.

We will consider the cubic phase gate \cite{Gottesman:2000di}
\begin{equation}
    \hat U_3 = e^{i\lambda\hat p^3},
    \label{eq:CubicGate}
\end{equation}
generated by $H=-\lambda\hat p^3$ acting for a unit amount of time. 
Since $\hat U_3$ is diagonal in momentum, it most naturally acts on the momentum Gaussian wavefunction
\begin{equation}
    \Phi(p) \equiv \langle p|\hat U_3|\Psi\rangle_G = e^{i\lambda p^3} \Phi_G(p)\, ,
\end{equation}
where $\Phi_G(p) \equiv \langle p|\Psi\rangle_G$ is the Gaussian state in momentum space.
The resulting third-order cumulants of $\hat U_3|\Psi\rangle_G$ follow from the commutation relation $[\hat p^3,\hat q]=-3i\hat p^2$
\begin{align}
    C_{qqq} &= \tfrac32\lambda - 18\lambda\,\sigma_{qp}^2 - 216\lambda^3\sigma_p^6, &
    C_{qqp} &= -12\lambda\,\sigma_{qp}\sigma_p^2, \nonumber\\
    C_{qpp} &= -6\lambda\,\sigma_p^4, &
    C_{ppp} &= 0 .
    \label{eq:CubicGateCumulants}
\end{align}
Defining the dimensionless skewness from the leading-order part of $C_{qqq}$,
\begin{equation}
    \gamma \equiv \frac{3\lambda}{2s^3} 
    \label{eq:GammaLambda}
\end{equation}
with $s^2=1/(4\sigma_p^2)$, we see that
\begin{equation}
    \frac{C_{qqq}}{s^3} = \gamma - 3\beta^2\gamma - \gamma^3\, ,
    \label{eq:CqqqGate}
\end{equation}
where $\beta \equiv \frac{\sigma_{qp}}{s\,\sigma_p}$.
For an uncorrelated Gaussian input state ($\beta=0$), to leading order in $\lambda$ the cubic phase gate state is given by the dimensionless skewness $C_{qqq} \simeq \gamma s^3$.

\begin{figure}[t]
\centering\includegraphics[width=\textwidth]{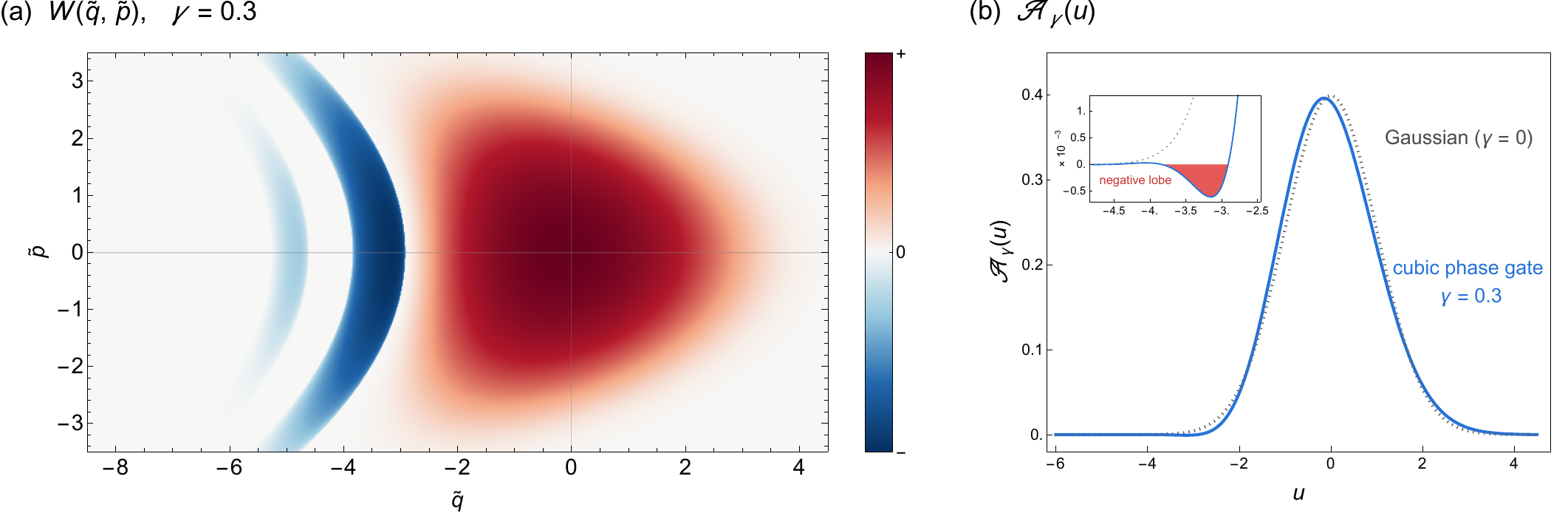}
\caption{(a) Wigner function \eqref{eq:CubicGateWigner} of an uncorrelated ($\sigma_{qp} = 0$) Gaussian state acted on by the cubic phase gate \eqref{eq:CubicGate} for $\gamma=0.3$, as a function of the dimensionless variables $\tilde q = q/s, \tilde p = p/\sigma_p$. The negative fringes lie along the parabolas $u=\tilde q+(\gamma/2)\tilde p^2=$ const.
(b) The negativity \eqref{eq:NegativityCubic} of the cubic phase gate Wigner function \eqref{eq:CubicGateWigner} arises from the scaled Airy function ${\mathcal A}(u)$ generated by the cubic phase gate (solid), compared here against
the Gaussian state with $\gamma = 0$ (dashed). The inset, magnified by $10^3$, shows
the negativity arising from the first negative lobe of the Airy function. Because of the structure of \eqref{eq:CubicGateWigner}, this plot is identical in form to Figure \ref{fig:AiryWigner}.}
\label{fig:CubicGateWigner}
\end{figure}

In the momentum representation the Wigner function is
\begin{align}
    W(q,p) & =\frac{1}{2\pi}\int dk\;e^{ikq}\,\Phi\big(p+\tfrac k2\big)\,\Phi^*\big(p-\tfrac k2\big)
    \label{eq:WignerMomentumRep1} \\
    & = \frac{1}{2\pi} \int dk\; e^{ik(q + 3\lambda p^2)} e^{\frac{i}{4} \lambda k^3} \Phi_G\left(p+\frac{k}{2}\right)\Phi_G^*\left(p-\frac{k}{2}\right) \label{eq:WignerMomentumRep2}
\end{align}
where we used $p_+^3 - p_-^3 = 3 p^2 k + k^3/4$ for $p_{\pm} = p\pm k/2$.
Using the inverse transform of the Gaussian momentum wavefunctions
\begin{align}
\Phi_G\big(p+\tfrac k2\big)\,\Phi_G^*\big(p-\tfrac k2\big)=\int dq'\,e^{-ikq'}\,W_G(q',p),
\end{align}
the Wigner function of the cubic gate can be rewritten as \cite{Ghose:2007qcy,Moore:2025qzc}
\begin{align}
    W(q,p) &= \int dq' W_G(q',p)\, \frac{1}{2\pi} \int dk\ e^{ik(q+3\lambda p^2-q')} e^{i\frac{\lambda}{4} k^3} \nonumber \\
    & = \frac{e^{-p^2/2\sigma_p^2}}{\sqrt{2\pi}\sigma_p}\ \frac{1}{s} {\mathcal A}_\gamma\left(\frac{Q'}{s}\right)
    \label{eq:CubicGateWigner}
\end{align}
where ${\mathcal A}_\gamma(u)$ is the same scaled Airy function  \eqref{eq:ScaledAiryFunction} from Section \ref{sec:Cumulant}, but now appearing as a function of the phase space variables $Q' = q - \frac{\sigma_{qp}}{\sigma_p^2} p + 3\lambda p^2$. Qualitatively, the cubic phase gate acts as classical shear, bending the strips of negativity seen in Figure \ref{fig:AiryWigner} into parabolas, as seen in Figure \ref{fig:CubicGateWigner}.

Transforming to the variables $(q,p) \rightarrow (Q',p)$, the negativity of \eqref{eq:CubicGateWigner} reduces to the negativity calculated earlier \eqref{eq:NegativityAiryWigner}
\begin{align}
    {\mathcal N} &= -2 \int dq\ dp\ \left[W(q,p)\right]_- = -2 \int dQ'\ dp\ \frac{e^{-p^2/2\sigma_p^2}}{\sqrt{2\pi}\sigma_p} \frac{1}{s} \left[{\mathcal A}_\gamma\left(\frac{Q'}{s}\right)\right]_- = -2 \int du\ \left[{\mathcal A}_\gamma(u)\right]_-\nonumber \\
    & \approx 2^{5/3} \gamma^{4/3} {\rm Ai}'(a_1) \exp\left[-\frac{1}{6\gamma^2} - \frac{|a_1|}{2^{1/3} \gamma^{2/3}}\right] \left[1 - 2^{2/3} |a_1| \gamma^{4/3} + {\mathcal O}(\gamma^2)\right]
    \label{eq:NegativityCubic}
\end{align}
This strongly suppressed physical Wigner negativity arises from a definite non-Gaussian deformation of a Gaussian state by a cubic phase gate that sources a cubic cumulant $\langle qqq\rangle$.

\bibliographystyle{utphysmodb}

\bibliography{refs}

\end{document}